\documentclass[pre,nofootinbib,superscriptaddress,showpacs,showkeys]{revtex4}

\usepackage{graphicx,epsfig}
\usepackage{amsfonts,amsmath,amssymb,amsthm,amscd}
\usepackage{color}

\begin{document}

\title{
Electromagnetic pulse reflection at self-generated plasma mirrors:
laser pulse shaping and high order harmonic generation.}

\author{S. S. Bulanov}
\affiliation{FOCUS center and Center for Ultrafast Optical Science,
University of Michigan, University of Michigan, Ann Arbor, USA}
\affiliation{Institute of Theoretical and Experimental Physics,
Moscow, Russia}

\author{A. Macchi}
\affiliation{polyLAB, CNR-INFM, University of Pisa, Pisa, Italy}

\author{A. Maksimchuk}
\affiliation{FOCUS center and Center for Ultrafast Optical Science,
University of Michigan, University of Michigan, Ann Arbor, USA}

\author{T. Matsuoka}
\affiliation{FOCUS center and Center for Ultrafast Optical Science,
University of Michigan, University of Michigan, Ann Arbor, USA}

\author{J. Nees}
\affiliation{FOCUS center and Center for Ultrafast Optical Science,
University of Michigan, University of Michigan, Ann Arbor, USA}

\author{F. Pegoraro}
\affiliation{Department of Physics, University of Pisa and CNISM,
Pisa, Italy}

\begin{abstract}
A thin layer of overdense plasma is created when an electromagnetic
pulse interacts with a rapidly ionizing thin foil. This layer will
reflect the incoming pulse, forming a so-called plasma mirror. A
simple realistic model based on paired kinetic and wave equations is
used to describe analytically the process of mirror formation and
the reflection and transmission of the incident pulse. The model
incorporates the exact description of the ionization process in the
foil and the polarization and conduction currents that follow. The
analytical description of the reflected and transmitted pulses as
well as their dependence on foil parameters, and initial pulse
amplitude and form are presented. Possible application and
effectiveness of this process to improve laser pulse contrast are
discussed. In the case of the linearly polarized incident pulse,
there harmonic generation occurs, that is absent in the case of the
circular polarization. The spectra of the reflected pulses for
different initial forms and amplitudes are studied.
\end{abstract}

 \maketitle

\section{Introduction}

The interaction of ultrashort, high-intensity laser pulses with
matter has lead to a new regime of nonlinear optics where the
optical properties of the medium undergo ultrafast changes due to
ionization produced by a laser pulse. In this context, by
ultrashort, high-intensity pulses we refer to pulses with typical
duration in the sub-picosecond regime, and whose intensity is high
enough to cause field ionization of matter within a very few optical
cycles, or even a sub-cycle interval. Therefore, by ultrafast
changes of, e.g., a change in the index of refraction we mean that a
latter varies significantly over a laser cycle, due to the rapid
increase of the free electron density due to field ionization. As a
consequence the pulse properties, such as envelope shape, spectrum
and duration, can be strongly modified.

A number of papers have addressed the problem of modeling the pulse
propagation in a fast ionizing medium (see e.g. \cite{ioniz1}) and
studied the possible applications of ionization effects for pulse
shaping and manipulation. The case of a high-density material, such
as a solid target, is of particular interest since the free electron
density produced by ionization can become quickly high enough to
make the plasma refractive index turn from real to imaginary values,
leading to self-reflection \cite{conejero,others}. This effect is
the physical basis for the so-called ``plasma mirror'' where plasma
production at the surface of a mirror is used experimentally to
obtain ``clean'' ultrashort pulses by removing the leading edge of
the low-intensity ``prepulse'' or ``pedestal'', which characterizes
femtosecond laser systems \cite{PM,PM1,PM2,PM3}, by letting it be
trasmitted through the mirror before the ionization--induced
transition to opacity occurs. The resulting improvement of the laser
pulse contrast (defined as the ratio between the prepulse energy and
the short pulse energy) is important for various applications of
current importance. These latter include, for example, the
acceleration of fast ions from solid targets (see \cite{ion_review}
and references therein) where avoiding plasma formation at the
target surface by the prepulse may enhance the ion source efficiency
\cite{neely}. Similarly, high-contrast pulse may optimize high
harmonic generation from solid target surfaces \cite{monot} Another
proposed application is the use of the combined effects of ultrafast
optical shuttering and harmonic generation to obtain a
sub-femtosecond pulse in transmission \cite{Bauer}.

In these application regimes a thin (i.e., having a thickness less
than an optical wavelength) foil target may be used in principle
because the thickness of the reflecting plasma layer is usually of
the order of the plasma skin depth. On the other hand the foil
thickness may be used as an additional control parameter for
applications. From the point of view of theoretical modeling a thin
foil target is of interest because it allows an analytical treatment
to some larger extent (see e.g. \cite{Bauer,BMP}), thus providing an
useful reference case for models and numerical simulations, and
possibly aiding further optimization of the plasma mirror
functionality. As a further reason of general interest, the
laser-plasma interaction dynamics in a thin foil includes peculiar
features that are not encountered either in underdense or overdense
plasmas \cite{Vshivkov,BMT}.

In the present paper we utilize a simple analytic model for a
dielectric foil ionization by an electromagnetic pulse. In order to
describe the circularly polarized pulse interaction with the foil we
use a kinetic equation paired with a wave equation. The effects of
ionization are incorporated into the source term in the kinetic
equation in the form of a differential ionization rate
\cite{keldysh,Popov}. The solution of the kinetic equation, the
momentum distribution function, enters the expression for the
conduction and polarization currents that emerge in the foil. We
point out differing behaviors between conduction and polarization
electric currents. While the first grows with the increase of
charged particle number, the second is proportional to the
ionization rate, which means that it will be most important in the
beginning of the ionization. In order to describe the field
evolution during the interaction with the foil we utilize the wave
equation with nonlinear boundary conditions, obtained in Ref.
\cite{BMP}. We show that as the pulse ionizes the foil its
reflectivity grows giving rise to the reflected pulse. The form of
the reflected and transmitted pulses depends strongly on the initial
amplitude of the pulse. We study the process of plasma mirror
formation under the action of a pulse with a gaussian envelope that
has a pedestal. There is a significant pedestal reduction after the
pulse reflection from the ionized foil. However there is a
dependence of the reduction efficiency on the initial contrast of
the pulse, which we study. We also consider the interaction of a
linearly polarized pulse with a thin dielectric foil. In this case
high order harmonics are generated due to the nonlinear dependence
of the ionization rate on the pulse amplitude. We study the high
order harmonic generation efficiency dependence on the initial
amplitude of the pulse and on the form of the pulse.

The paper is organized as follows. In section 2 we review the
solution of the Maxwell equations with nonlinear boundary
conditions. In section 3 we carry out the kinetic description of the
currents that enter the wave equation. In section 4 we consider an
analytical solution of the simplified wave equation. The results of
the numerical solution in the case of a circularly polarized initial
pulse are presented in section 5. In section 6 a linearly polarized
pulse interaction with a foil is studied numerically. We summarize
conclusions in section 7.

\section{Wave equation with nonlinear boundary conditions}\label{WE}

In this section we review the solution of the wave equation with
nonlinear boundary conditions as it was presented in Ref.
\cite{BMP}. We consider the case of  a circularly polarized plane
wave impinging at normal incidence on a hydrogen-like foil, which is
located in the plane $x=0$. The interaction of the pulse with the
foil is taken to be uniform along the foil surface, which is
justified by the fact that plasma mirrors operate in the regime in
which the laser pulse is weakly focused on a mirror with a spot size
of hundreds of wavelengths \cite{PM,PM1,PM2,PM3}. Relativistic
effects of the ionization of the foil are neglected, which is valid
if the magnetic field contribution to the ionization probability as
well as to the electron motion is small. The density of the foil is
taken to be high enough to ensure that the plasma density is not
restrained by the fact that the number of neutral atoms might be
exhausted. Also it is assumed that the electrons that emerge in the
foil during the ionization remain localized within the foil due to
the charge separation field produced in the presence of immobile
heavy ions. So, the current is proportional to $\delta(x)$, where
$\delta(x)$ is Dirac delta function. The wave equation for the
vector potential $\mathbf{A}(x,t)$ is
\begin{equation} \label{wave}
\partial_{tt}\mathbf{A}-c^2\partial_{xx}\mathbf{A}=4\pi \,c\, \delta(x)
\mathbf{J}(\mathbf{A})+\delta^\prime(t)\mathbf{A}(x,0)+\delta(t)\dot{\mathbf{A}}
(x,0).
\end{equation}
The first term on the RHS of Eq. (\ref{wave}) describes the electric
current, localized in the foil, which is a functional of the vector
potential taken at $x=0$. The last two terms are equivalent to the
initial conditions $\mathbf{A}(x,0)=\mathbf{A}_0(x)$, $\partial_t
\mathbf{A}(x,0)=\dot{\mathbf{A}}_0(x)$, where $\mathbf{A}_0(x)$ and
$\dot{\mathbf{A}}_0(x)$ define the incident electromagnetic pulse.

The solution of Eq. (\ref{wave}) can be presented as \cite{BMP}:
\begin{equation} \label{sol wave}
\mathbf{A}(x,t)=\mathbf{A}_0(x,t)+2\pi\int\limits_0^{t-|x|/c}
\mathbf{J}\left(\mathbf{A}(0,\tau)\right)d\tau
\end{equation}
By differentiating this equation with respect to time or spatial
coordinate $x$ we obtain the explicit form of the electric or
magnetic field respectively as functions of the incident electric
field $\mathbf{E}_0(x,t)$ and $\mathbf{A}(0,t)$, which can be found
from Eq. (\ref{sol wave}) taken at $x=0$.
\begin{equation} \label{E}
\mathbf{E}(x,t)=\mathbf{E}_0(x,t)-\frac{2\pi}{c}\,
\mathbf{J}\left(\mathbf{A}(0,t-|x|/c)\right),
\end{equation}
\begin{equation} \label{B}
\mathbf{B}(x,t)=\mathbf{B}_0(x,t)+\frac{2\pi}{c}\, \mathbf{e}_x
\times \mathbf{J}\left(\mathbf{A}(0,t-|x|/c)\right) \text{sign}(x),
\end{equation}
where $\mathbf{e}_x$ is the unit vector along the direction of wave
propagation. In order to solve these equations one should determine
the expression for the current.

\section{Kinetic description of the current}\label{KD}

The behavior of the electrons in the foil in the presence of the
homogeneous time-varying electric field can be described by the
kinetic equation of the following form
\begin{equation} \label{kinetic}
\frac{\partial f(\mathbf{p},t)}{\partial
t}-e\mathbf{E}(t)\frac{\partial f(\mathbf{p},t)}{\partial
\mathbf{p}}=q(\mathbf{p},\mathbf{E},t),
\end{equation}
where the distribution function $f(\mathbf{p},t)$ is normalized in
such a way that $\int f(\mathbf{p},t) d^3p$ gives the number of
electrons per unit volume. (Similar approaches to field ionization
modeling can be found, e.g., in Ref.\cite{mulser,cornolti}). Since
we are considering the nonrelativistic regime, we can neglect the
effects of the incident magnetic field, which are suppressed by the
factor $v/c$, in the motion of electrons as well as in the
ionization. The source term, $q(\mathbf{p},\mathbf{E},t)$, is
connected with ionization of the foil by the incident
electromagnetic pulse in such a way that $\int
q(\mathbf{p},\mathbf{E},t) d^3p$ gives the total ionization
probability $W=8 n_a\omega K_0\kappa^3\mathcal{E}^{-1}
\exp\left[-2\kappa^3/3\mathcal{E}\right]$ of the unit volume in unit
time (the time is measured in units of $\hbar^3/me^4$). It is
determined by the differential probability of ionization by a
circularly polarized electric field \cite{Popov}
\begin{equation}
q=\frac{4\kappa^4}{\pi^2}\frac{n_a\omega K_0}{\mathcal{E}^2}
\exp\left[-\frac{2\kappa^3}{3\mathcal{E}}\right]
\exp\left\{-\frac{\kappa}{\mathcal{E}}
\left[\left(\frac{p_\perp}{mc}-\frac{\mathcal{E}}{\omega}
\frac{(4\pi\alpha)^2
c}{a_B}\right)^{2}+\left(\frac{p_x}{mc}\right)^{2}\right]\right\},
\end{equation}
where $\omega$ is the frequency of the incident electric field,
$\mathcal{E}=E/E_a$ is the normalized amplitude of the electric
field, $E_a=m^2e^5/\hbar^4=5.14\cdot 10^9$ V/cm is characteristic
atomic field, $\kappa=\sqrt{I/I_H}$, $I$ is the ionization potential
and $I_H=me^4/2\hbar^2=13.6$ eV is the ionization potential of the
hydrogen atom, $K_0=I/\hbar\omega$, $n_a$ is the density of neutral
atoms, $a_B=\hbar^2/me^2$ is Bohr radius, and $\alpha=1/137$ is
fine-structure constant. The component of electron momentum
perpendicular to the direction of pulse propagation is denoted by
$p_\perp$, and along the direction of pulse propagation is $p_x$.
The momentum distribution of produced electrons is very narrow with
maximum at $p_x=0$, $p_\perp/mc=(\mathcal{E}/\omega)(16\pi^2\alpha
c/a_B)$ \cite{Popruzhenko,MurPopov}. This means that electrons
preferably emerge from under the barrier with momentum perpendicular
to the instantaneous direction of the electric field and equal to
that of a free electron acquires in such a field. Since the momentum
distribution is very narrow, we can use delta functions instead of
exponents in order to simplify the calculations:
\begin{equation} \label{source}
q=\frac{8 \kappa^3 K_0 n_a\omega}{\mathcal{E}}
\exp\left[-\frac{2\kappa^3}{3\mathcal{E}}\right] \delta
\left(\frac{p_\perp}{mc}-\frac{\mathcal{E}}{\omega}
\frac{(4\pi\alpha)^2 c}{a_B}\right)\delta(p_x).
\end{equation}
We should also note that in the expression for the source term the
absolute value of the field enters.

We solve Eq.(\ref{kinetic}) by integrating it  along the particle
characteristics. The equations for the characteristics are
\begin{equation}
{p}_{0} = {p}_{0}(0), \qquad \frac{dp_{\perp }}{dt}=-eE,\qquad
\frac{df}{dt}=q.
\end{equation}
Introducing the function
$\mathbf{A}(t)=-\int\limits_{0}^{t}\mathbf{E}ds$, we obtain
\begin{equation}
\mathbf{p}_{\perp}=-e\int\limits_{0}^{t}\mathbf{E}ds+\mathbf{p}_{\perp
0}\quad {\rm i. e.,} \quad \mathbf{p}_{\perp }-e\mathbf{A}(t)=
\mathbf{p}_{\perp 0}.
\end{equation}
As a result the distribution function is given by the  following
functional
\begin{equation}
f=f_0[p_x, \mathbf{p}_{\perp
}-e\mathbf{A}(t)]+\int\limits_{0}^{t}q\{\mathbf{p}_{\perp
}-e[\mathbf{A}(t)-\mathbf{A}(t^{\prime })],t^{\prime }\}dt^{\prime
}, \label{f}
\end{equation}
where $f_{0}(p_x, \mathbf{p}_{\perp })$ is the distribution function
of the initial electrons before the passage of the electromagnetic
wave. In our case $f_{0}=0$, since the foil was not ionized
initially. The appearance of electrons due to the ionization of the
foil gives rise to conduction and polarization electric currents.
The conduction electric current is due to the electron motion in the
field of the incident pulse, while the polarization electric current
is due to the ionization process, since when the neutral atom is
ionized an electric dipole is created.

In order to determine the form of the polarization electric current
we write down the second moment of the kinetic equation
\begin{equation}
\label{2moment}
\begin{tabular}{rc}
$\sum\limits_\alpha\int\left[(m_\alpha^2c^4+p^2c^2)^{1/2}-m_\alpha
c^2\right] \left\{\frac{\partial f_\alpha(\mathbf{p},t)}{\partial t}
+e_\alpha \mathbf{E}(t) \frac{\partial
f_\alpha(\mathbf{p},t)}{\partial \mathbf{p}}\right\}d^3p$ \\&\\
$=\sum\limits_\alpha\int\left[(m_\alpha^2c^4+p^2c^2)^{1/2}-m_\alpha
c^2\right]q_\alpha (\mathbf{E},\mathbf{p},t) d^3p,$
\end{tabular}
\end{equation}
which we rewrite as
\begin{equation} \label{K}
\frac{\partial K}{\partial t}-\mathbf{j}_{cond}\mathbf{E}=\Sigma
-\sum\limits_\alpha m_\alpha\frac{\partial n_\alpha}{\partial t},
~~~\mathbf{j}_{cond}=e\int f\frac{\mathbf{p}_\perp
c}{\sqrt{m^2c^4+\mathbf{p}^2 c^2}} d^3p,
\end{equation}
where $K=\sum\limits_\alpha\int\left[(m_\alpha^2c^4+p^2c^2)^{1/2}
-m_\alpha c^2\right]f_\alpha d^3p$ is the kinetic energy and
$\Sigma=\sum\limits_\alpha\int(m_\alpha^2 c^4+p^2 c^2)^{1/2}q_\alpha
(\mathbf{E},\mathbf{p},t) d^3p$. Since in the foil not only
electrons and ions are present but also neutral atoms, from which by
the process of ionization the plasma is created, $\partial
n_e/\partial t=\partial n_i/\partial t=-\partial n_a/\partial t$.
Here indexes $e$, $i$, and $a$ stand for electron, ion, and atom
respectively. Then
\begin{equation}
\sum\limits_\alpha m_\alpha c^2\frac{\partial n_\alpha}{\partial
t}=\left(m_e+m_i-m_a\right)c^2 \frac{\partial n_e}{\partial
t}=I\frac{\partial n_e}{\partial t},
\end{equation}
where $I=\left(m_e+m_i-m_a\right) c^2$, and is usually referred to
as a mass defect or ionization potential, which is due to the fact
that some electrons and ions are bound in atoms. In order to ionize
an atom some amount of energy should be spent. So we rewrite Eq.
(\ref{K}) in the following form
\begin{equation}
\frac{\partial}{\partial t}\left(K+In_e\right)=
\mathbf{j}_{cond}\mathbf{E}+\Sigma.
\end{equation}
The energy balance equation for the circularly polarized
electromagnetic wave is
\begin{equation}
\frac{\partial}{\partial t}\left(\frac{\mathbf{E}^2}{4\pi}\right)=
-\mathbf{j}_{cond}\mathbf{E}-\mathbf{j}_{pol}\mathbf{E}.
\end{equation}
Here we have taken into account that no external electric current is
present in our case. Adding these two equations we obtain the energy
balance equation for the system of electromagnetic wave and
particles
\begin{equation}
\frac{\partial}{\partial t}\left(\frac{\mathbf{E}^2}{4\pi}+ K+I
n_e\right)= -\mathbf{j}_{pol}\mathbf{E}+\Sigma.
\end{equation}
The RHS of this equation should be equal to zero, because of the
energy conservation law. Since we are considering a nonrelativistic
case, we can neglect $p_\alpha$ with respect to $m_\alpha$. Then the
expression for polarization current will take the following form
\begin{equation} \label{jpol}
\mathbf{j}_{pol}=
\frac{\mathbf{E}(t)}{|\mathbf{E}(t)|^2} I\frac{\partial n_e}{\partial t}.
\end{equation}
To our knowledge the polarization current was first introduced in
Ref.\cite{rae-burnett} while an extended discussion is given in
Ref.\cite{mulser}. The expression given in Ref.\cite{mulser} is
found to be identical to our formula (\ref{jpol}) when one realizes
that the ``ejection energy'' of freed electrons ${\cal E}_I$ has
been included in $I$. For our aims one may assume ${\cal E}_I=0$,
i.e. we assume free electrons appear with zero energy, as an
acceptable simplifying assumption; in such a case $I$ represents in
practice the standard ionization potential. In the more general case
$I$ becomes a smooth function of the electric field.

It can be seen from Eq. (\ref{jpol}) that the polarization electric
current is along the direction of the electric field and is
proportional to the ionization potential and to the ionization rate.
Due to this fact we expect the contribution of the polarization
electric current to be rather small. However, if we consider the
ionization of heavy multicharged ions (where the binding energy of
low lying levels is comparable to the electron rest mass) by an
intense electromagnetic pulse, the polarization electric current can
play an important role. It is similar to the case of
electron-positron pair production by intense electromagnetic wave in
plasma \cite{BFP}, where the polarization electric current is
proportional to $2m_e$. In that case the polarization electric
current totally changes the behavior of the wave, forcing the vector
potential to oscillate around some nonzero value. Also the
components of the electric field begin to oscillate with different
frequencies. So in the process where the backreaction of the
produced particles on the properties of the wave that created them
takes place, the polarization electric current should be taken into
account. In the case of a very sharp electromagnetic pulse front,
consider, for example, the interaction of high contrast and very
intense pulse with the foil, the ionization rate is very high, and
the polarization electric current will be significant, since the
conduction electric current will not be able to develop large value
due to the absence of electrons and ions in the first moments of
evolution.

Using the explicit expression for the distribution function
(\ref{f}) and the source term (\ref{source}), and substituting
(\ref{K},\ref{jpol}) into Eq. (\ref{E}), we obtain:
\begin{eqnarray}
\mathbf{E}(x,t)=\mathbf{E}_0(x,t)-\frac{2\pi}{c}\frac{e^2 l}{m}
\mathbf{A}(0,t-|x|/c)\int\limits_0^{t-|x|/c}\frac{8\kappa^3 K_0
n_a\omega}{\mathcal{E}}
\exp\left[-\frac{2\kappa^3}{3\mathcal{E}}\right] dt^\prime \nonumber
\\ -\frac{2\pi}{c}\frac{\mathbf{E}(0,t-|x|/c)}{|\mathbf{E}(0,t-|x|/c)|^2}
\frac{8 I\,l\kappa^3 K_0 n_a\omega}{\mathcal{E}}
\exp\left[-\frac{2\kappa^3}{3\mathcal{E}}\right].
\end{eqnarray}
We can rewrite the equation for the electric field in terms of
dimensionless field and vector potential, $\mathbf{\eta}=
e\mathbf{E}/m c\omega_0$, $\mathbf{a}=e\mathbf{A}/m c^2$, where
$\omega_0$ is the frequency of the incident wave. The density of the
plasma, that emerges due to the ionization, is taken in the units of
$n_0=m\omega_{pe}^2/4\pi e^2$, the critical plasma density,
corresponding to the frequency of the initial pulse with time in the
units of inverse frequency, $\omega_0$. The equation that governs
the behavior of the dimensionless electric field is
\begin{equation} \label{field}
\mathbf{\eta}(x,t)=\mathbf{\eta}_0(x,t)-\epsilon_p
\mathbf{a}(0,t-|x/c|) n(0,t-|x/c|) - \epsilon_p\frac{I}{mc^2}
\frac{\mathbf{\eta}(0,t-|x/c|)}{|\mathbf{\eta}(0,t-|x/c|)|^2}
\dot{n}(0,t-|x/c|),
\end{equation}
where
\begin{equation}
n(t)=\frac{n_a}{n_0}\int \limits_0^t \frac{8\kappa^3K_0\eta_a}
{|\mathbf{\eta}(t^\prime)|}
\exp\left[-\frac{2\kappa^3\eta_a}{3|\mathbf{\eta}(t^\prime)|}\right]
dt^\prime ~~~\text{and}~~~
\dot{n}(t)=\frac{n_a}{n_0}\frac{8\kappa^3K_0\eta_a}
{|\mathbf{\eta}(t)|}
\exp\left[-\frac{2\kappa^3\eta_a}{3|\mathbf{\eta}(t)|}\right],
\end{equation}
The normalized atomic field is denoted as $\eta_a= e E_a/m\omega c$
and the normalized density of the foil, $\epsilon_p=\omega_{pe}^2
l/2\omega_0 c$, is a crucial parameter of the laser pulse
interaction with a thin plasma layer introduced in Ref.
\cite{Vshivkov}. It plays an important role not only in the process
of electromagnetic pulse reflection/transmission at the thin foil
\cite{BMP}, but also in the process of ion acceleration
\cite{Vshivkov}, when an intense laser pulse interacts with the thin
foil, as well as in the process of high order harmonic generation in
the relativistic regime of laser-foil interaction \cite{Pirozhkov}.

Equation (\ref{field}) demonstrates that the conduction electric
current is proportional to the number of electrons (ions), i.e. to
plasma density, while the polarization electric current is
proportional to the ionization rate, which makes it especially
important in the first moments of intense ionization, should it
occur. This distinction is due to the fact that the conduction
electric current is connected with the motion of charged particles,
while the polarization electric current is due to dipole creation
through the ionization of neutral atoms.

Due to the nonlinear dependence of the current on the electric field
amplitude the above equation is difficult to solve analytically.

\section{Model case}

In order to study the properties of Eq. (\ref{field}) we consider a
simplified model case. Let us assume that $n(t)$ and thus
$\dot{n}(t)$ are given functions of time. Then the equation
(\ref{field}), governing the evolution of the electric field, can be
rewritten at the foil ($x=0$) in the following form
\begin{equation} \label{field_S}
\mathbf{\eta}=\mathbf{\eta}_0-
N\mathbf{a}+\frac{\mathbf{\eta}}{|\mathbf{\eta}|^2}N_1,
~~~\text{where}~~N=\epsilon_p n(t)~~\text{and}
~~N_1=\epsilon_p\frac{I}{mc^2}\frac{\partial n(t)}{\partial t}.
\end{equation}
The initial electromagnetic wave is
\begin{equation}
\mathbf{\eta}_0=\eta_{in} e^{it}.
\end{equation}
We shall look for the solution of Eq. (\ref{field_S}) in the
following form
\begin{equation}
\mathbf{\eta}=\eta e^{i(t+\varphi)}.
\end{equation}
After substituting the initial field and the solution into Eq.
(\ref{field_S}) we obtain a system of algebraic equations for the
amplitude and the phase of the transmitted wave:
\begin{equation} \label{system}
\begin{array}{l}
(\eta^2+N_1)\cos\varphi=\eta\eta_{in}-\eta^2 N \sin\varphi \\
\\
(\eta^2+N_1)\sin\varphi=\eta^2 N \cos\varphi.
\end{array}
\end{equation}
Using the second equation we get the following expression for
$\varphi$:
\begin{equation}\label{phase}
\varphi=\arctan\frac{\eta^2 N}{\eta^2+N_1}.
\end{equation}
Substituting this expression into the first equation of the system
(\ref{system}), we find a forth order algebraic equation for $\eta$:
\begin{equation}
\eta^{4}(1+N^2)+\eta^2(2N_1-\eta_{in})+N_1^2=0.
\end{equation}
The solutions of this equation are
\begin{equation}
\eta=\pm\left[\frac{-(2N_1-\eta_{in}^2)+\sqrt{(2N_1-\eta_{in}^2)^2-4(1+N^2)N_1^2}}
{2(1+N^2)}\right]^{1/2}.
\end{equation}
Two other roots of forth order equation were omitted since $\eta$
should be real.

Let us consider the moment of time, when there is no longer any
ionization, because all the atoms that could be ionized by the pulse
have been ionized. The plasma density reached its maximum and the
main part of the incoming radiation is reflected by plasma mirror.
In this case $N_1$ has $\partial n(t)/\partial t=0$, and
\begin{equation} \label{tran}
\eta=\pm\frac{\eta_{in}}{\sqrt{1+N^2}}.
\end{equation}
This result states that even for plasma density much greater than
critical there is a transmission of the incoming pulse through the
ionized foil. However this transmission is suppressed by increased
plasma density.

Let us investigate the phase behavior. Using Eq. (\ref{phase}), we
find
\begin{equation}
\varphi=\arctan\frac{\eta^2 N}{\eta^2+N_1}\approx \arctan N.
\end{equation}
The phase variation with time is
\begin{equation}
\frac{\partial \varphi}{\partial t}=\frac{\dot{N}}{1+N^2}>0.
\end{equation}
So there is a blue shift in transmitted pulse frequency, while the
density of plasma increases (see also \cite{koga}). However in the
region with $N=const$ and $N_1=0$ there is no frequency shift and
the radiation is transmitted without any frequency change. This fact
should lead to the steepening of the transmitted pulse front and the
rear should be of the same (half)width as the initial pulse. This
can be seen from the results of numerical calculations, presented in
the next section of this paper. So here unlike the case of the wave
propagation in the ionized medium, the blue shift is in effect only
when the ionization occurs. The frequency up-shift is connected with
the rate of plasma density increase.

\section{The results of numerical solution}\label{L1}

In this section we present the results of numerical solution of Eq.
(\ref{field}) at the foil, $x=0$. We consider two cases of initial
pulse form. First, the case of the pulse with gaussian envelope and
gaussian pedestal, the maxima of the pulse and the pedestal
temporaly coincide,
\begin{equation} \label{pedestal}
E(0,t)=E_0\left\{p
\exp\left[-\left(\frac{t-\tau/2}{\tau/2}\right)^2\right] +(1-p)
\exp\left[-\left(\frac{t-\tau/2}{\tau/20}\right)^2\right] \right\}
e^{i\omega_0 t},
\end{equation}
where $p$ is the ratio of pedestal amplitude to the main pulse
amplitude and $\tau=40~T$ is the pulse duration, $T=2\pi/\omega_0$.
Second, the case of
\begin{equation} \label{sin}
E(0,t)=E_0[\sin (\pi t/\tau)]^2 \exp[i\omega_0 t],
\end{equation}
We consider time interval $(0,\tau)$ to ensure that only one pulse
is taken into account. This form of the pulse allows us to
investigate the fronts of the reflected and transmitted pulses,
since at $t=0$ the initial field is equal to zero.

\begin{figure}[ht]
\begin{tabular}{ccc}
\epsfxsize5cm\epsffile{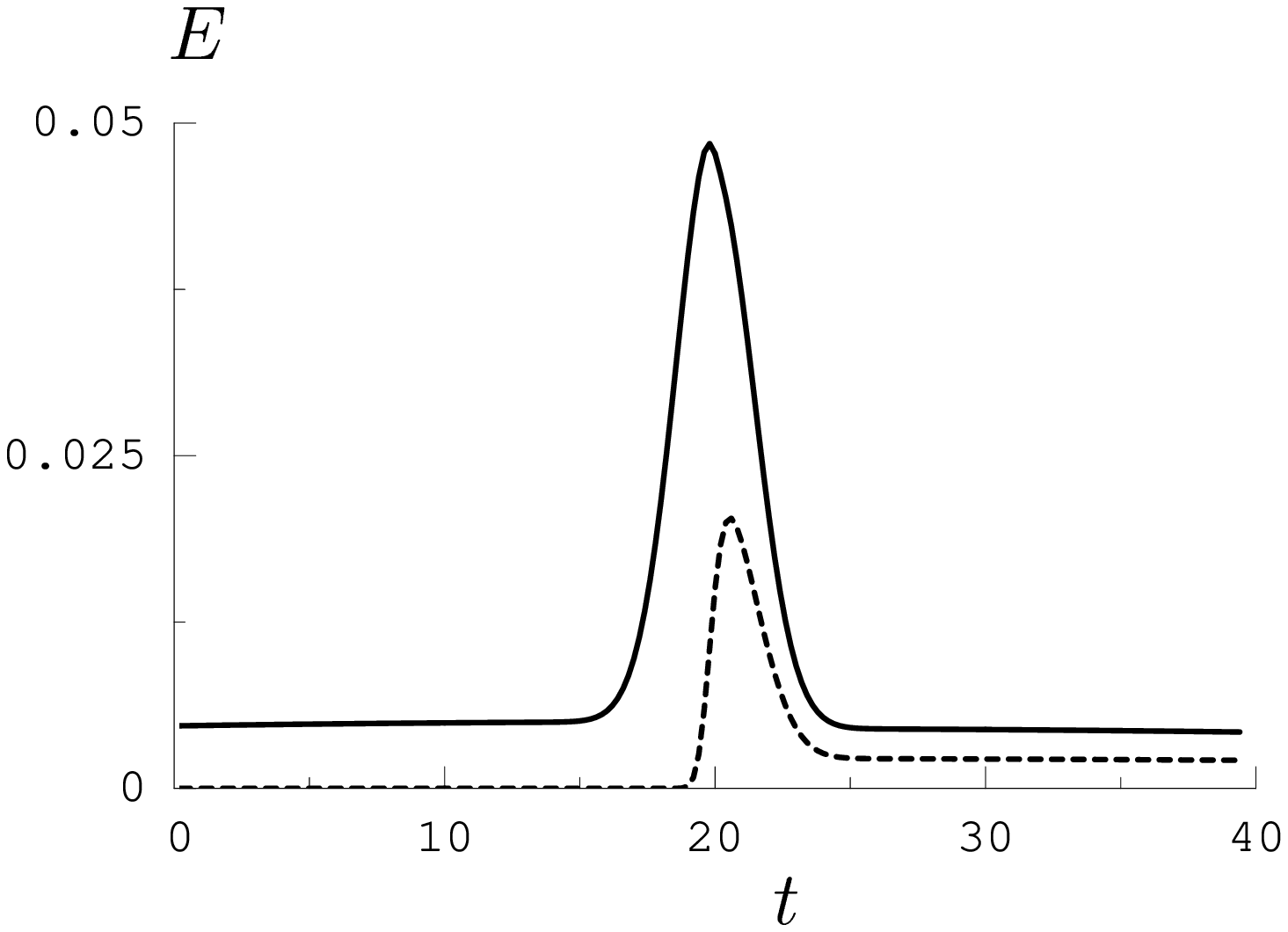} &
\epsfxsize5cm\epsffile{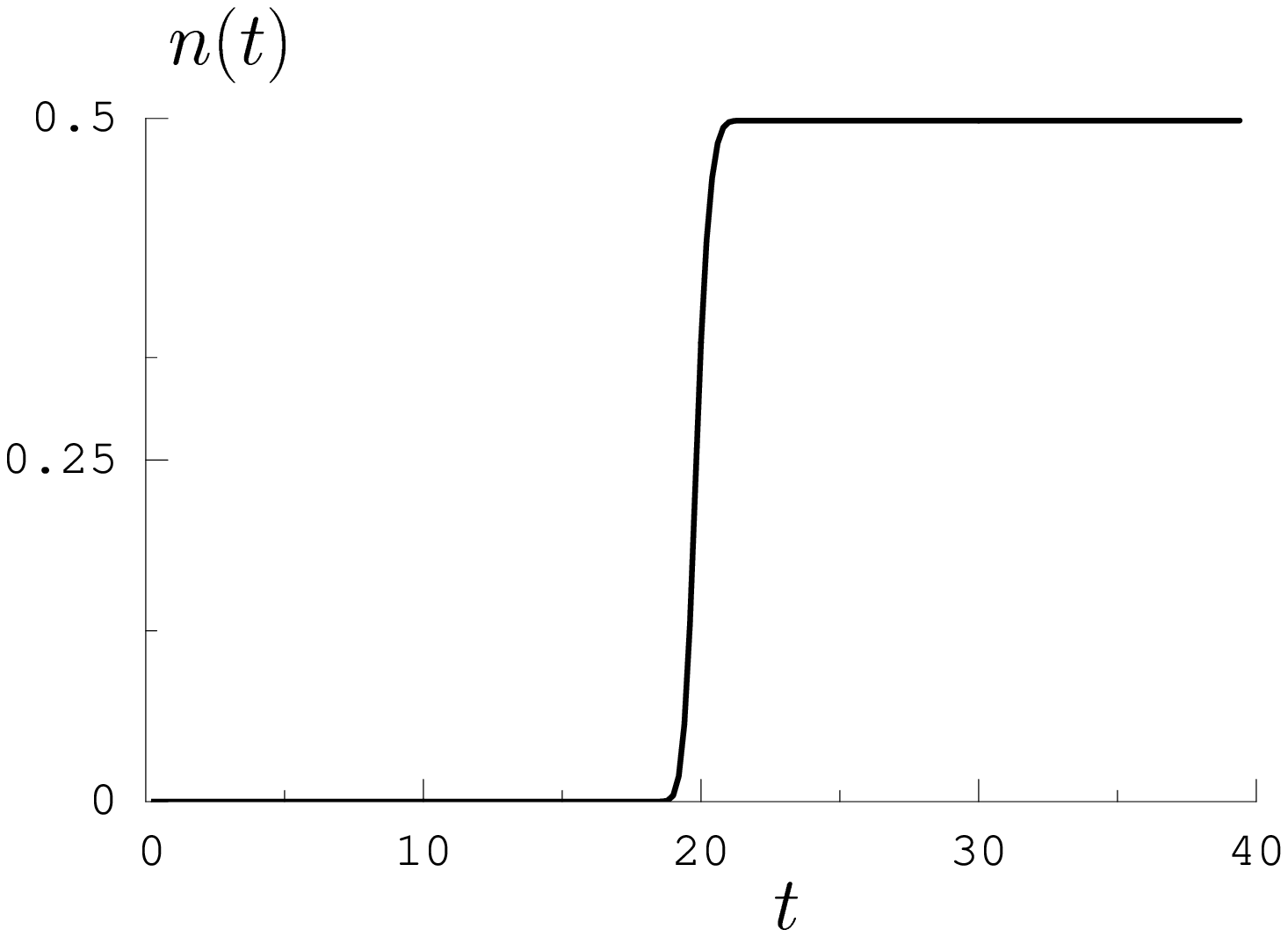} &
\epsfxsize5cm\epsffile{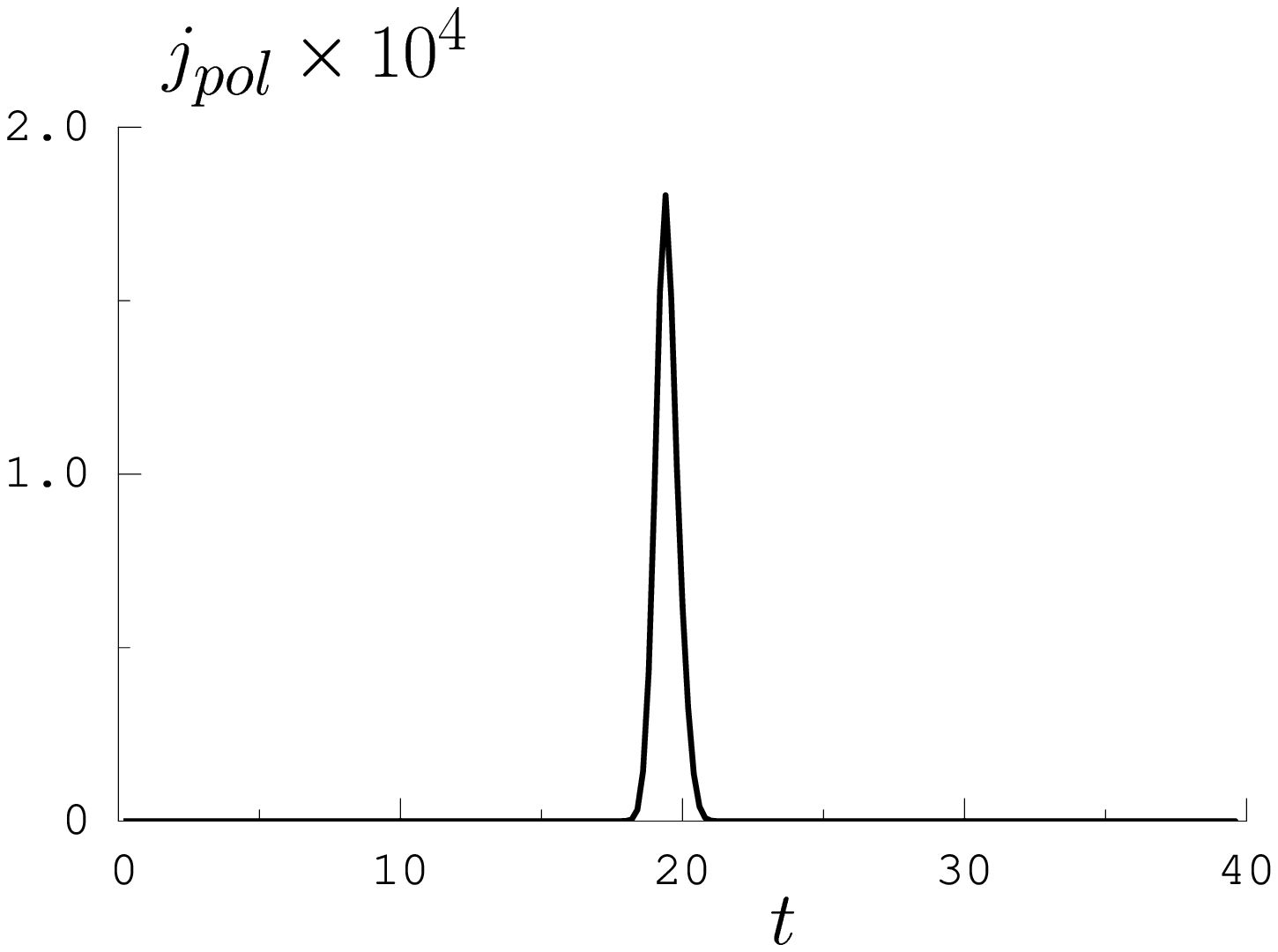}\\
\epsfxsize5cm\epsffile{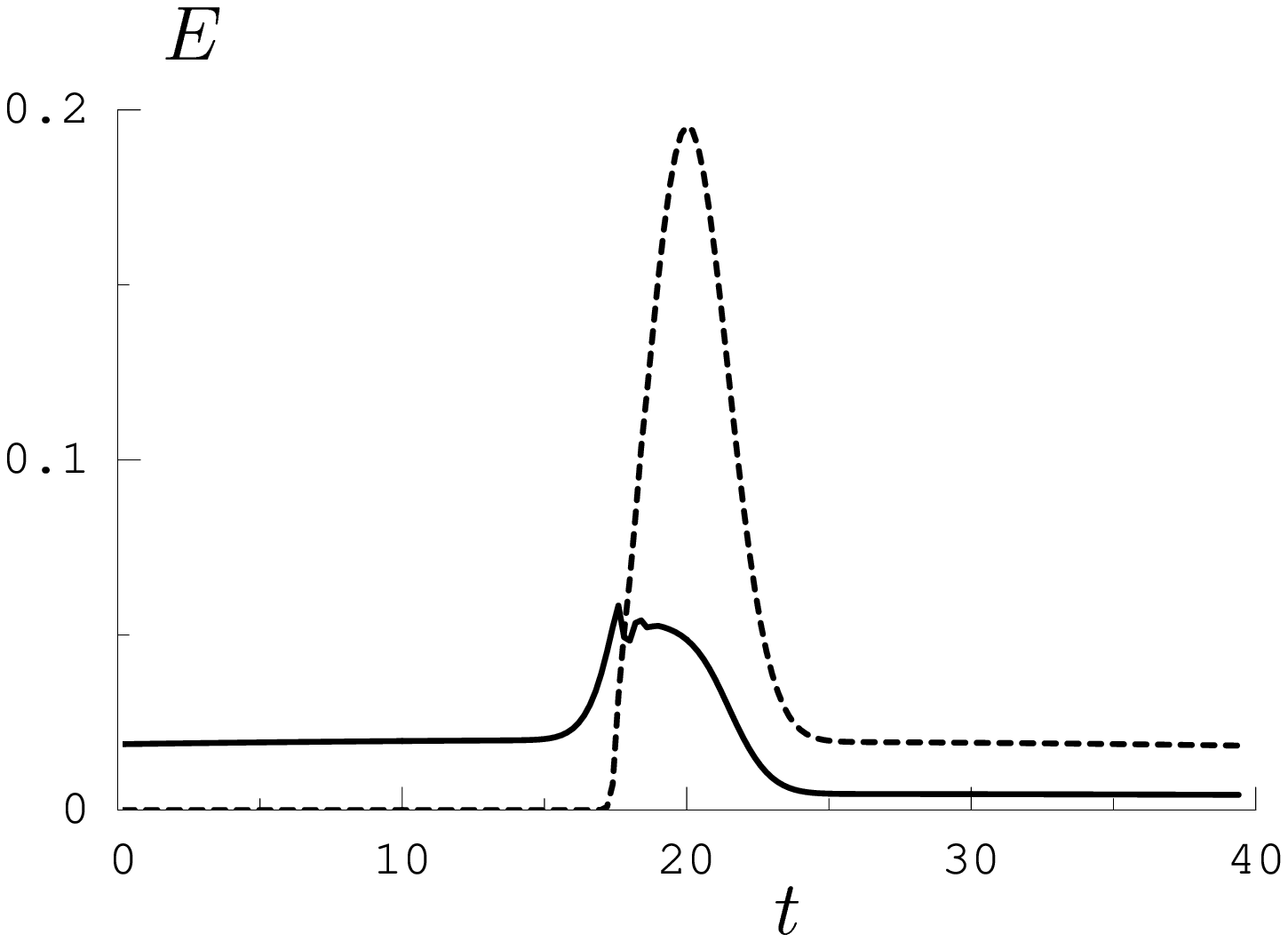} &
\epsfxsize5cm\epsffile{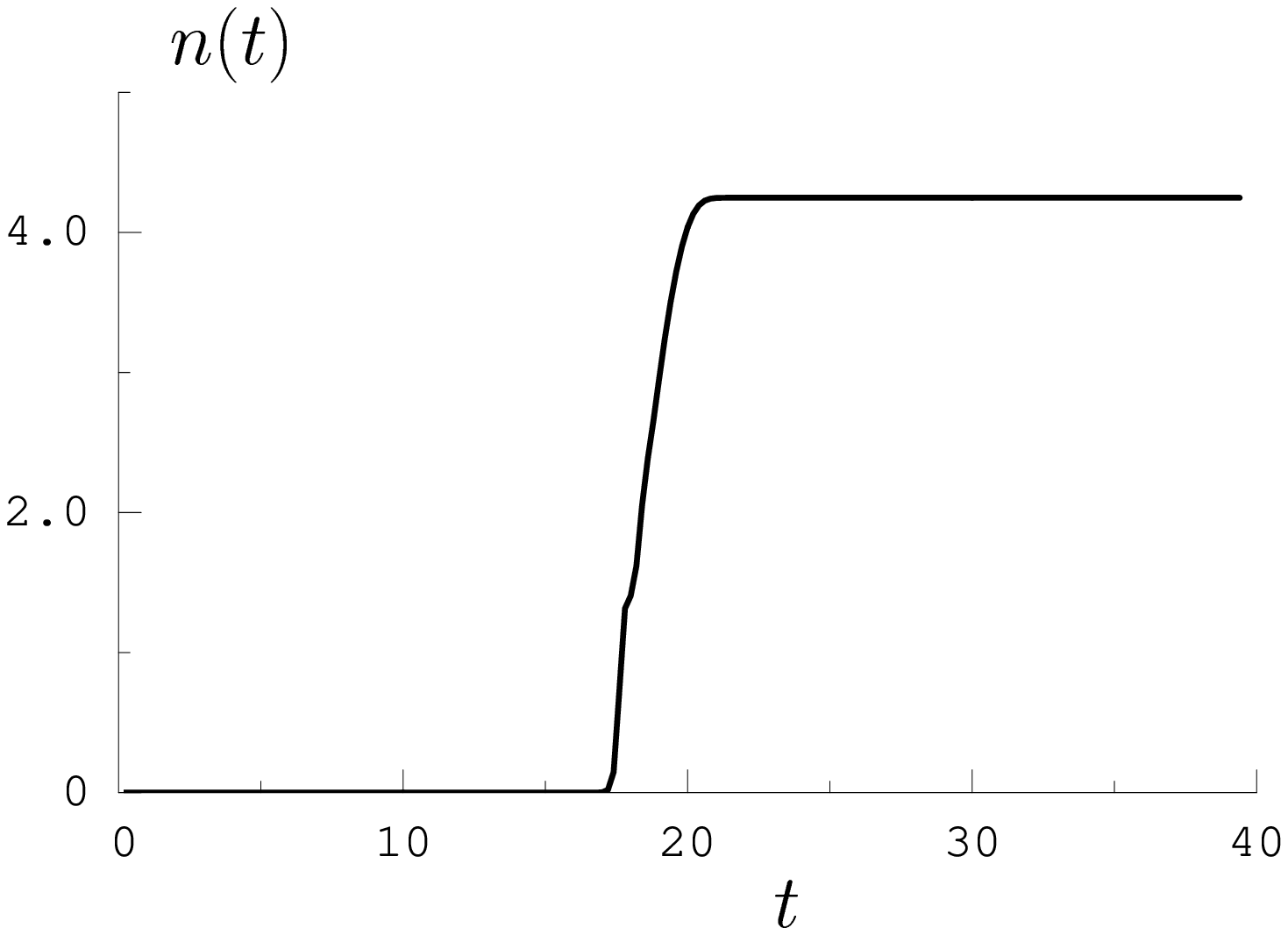} &
\epsfxsize5cm\epsffile{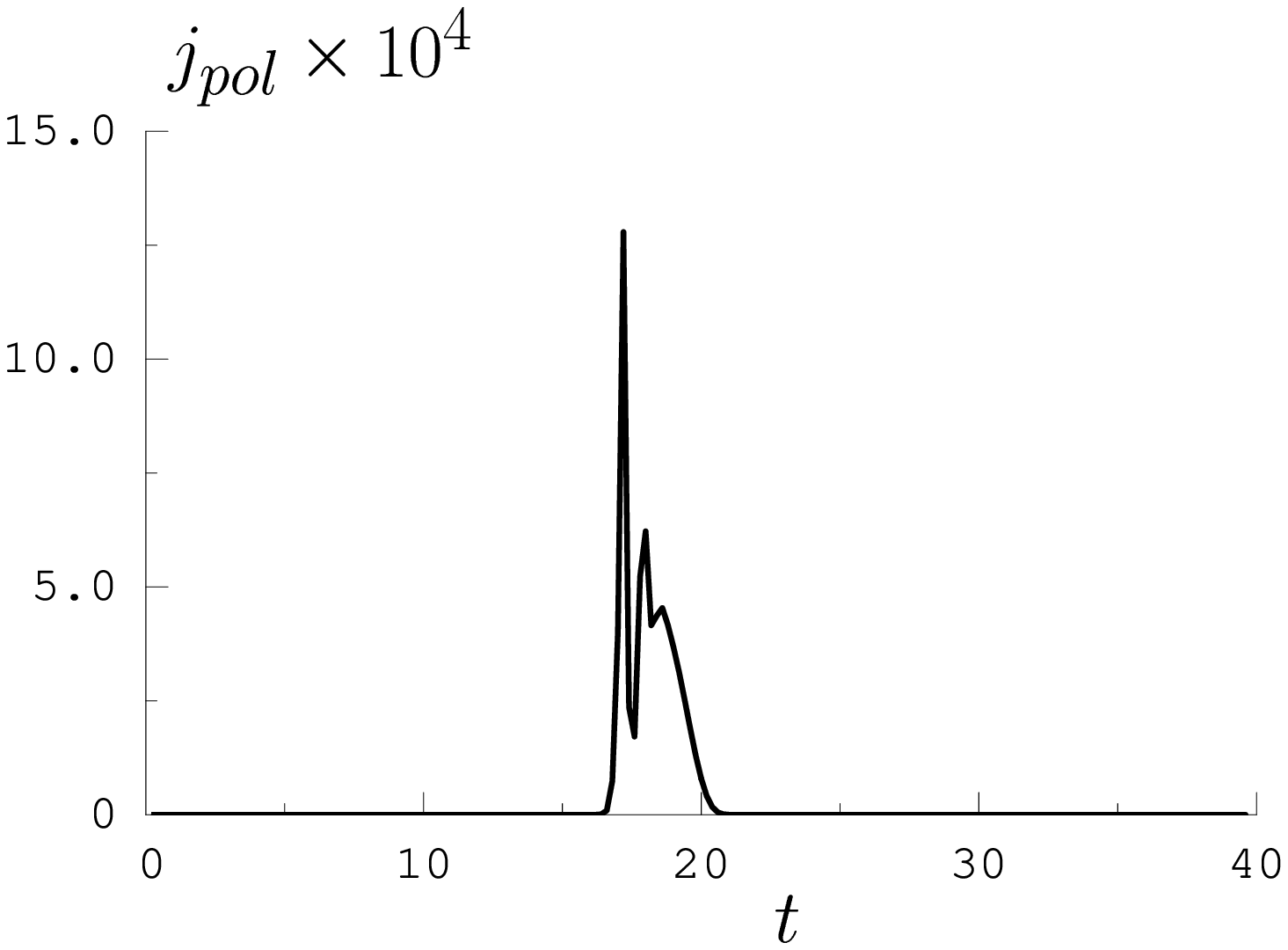}\\ a) & b) & c)\\
\end{tabular}
\caption{Evolution of the
reflected (dotted curve) and transmitted pulse electric field (a),
plasma density (b) and polarization current (c) for initial pulse of
the form (\ref{pedestal}). Time is measured in units of wave period
($2\pi/\omega$), field is measured in units of $mc\omega/e$, density
is measured in units of critical density, $n_{cr}$, and currents are
dimensionless. The parameter $\epsilon_p K_0=4\times 10^3$, $x=0$.
The first row corresponds to the initial pulse amplitude equal to
$\eta_0^{max}=0.05$, the second row to $\eta_0^{max}=0.2$.}
\end{figure}

In Figs. 1 and 2 we present the results of numerical solution of Eq.
(\ref{field}) with initial pulse having form (\ref{pedestal}) and
(\ref{sin}) respectively. In Fig. 1 maximal electric field is
$\eta_0^{max}=0.05$ (first row) and $\eta_0^{max}=0.2$ (second row).
The amplitudes of the reflected and transmitted pulses are shown in
Fig. 1a. The plasma density evolution which is closely connected
with the conduction current is shown in Fig. 1b. The behavior of
polarization current is presented in Fig. 1c. In the case of
$\eta_0^{max}=0.05$ (Fig. 1, first row) the incoming pulse is not
intense enough to sufficiently ionize the foil, the resulting plasma
density is 0.5 of critical. Due to this fact the reflectivity of the
of the ionized foil is low, and the reflected pulse is much lower
than the transmitted one. However it can clearly be seen that the
pedestal is transmitted through the foil. The reflected pulse starts
to differ from zero only when the main part of the initial pulse
arrives at the foil.

\begin{figure}[ht]
\begin{tabular}{ccc}
\epsfxsize5cm\epsffile{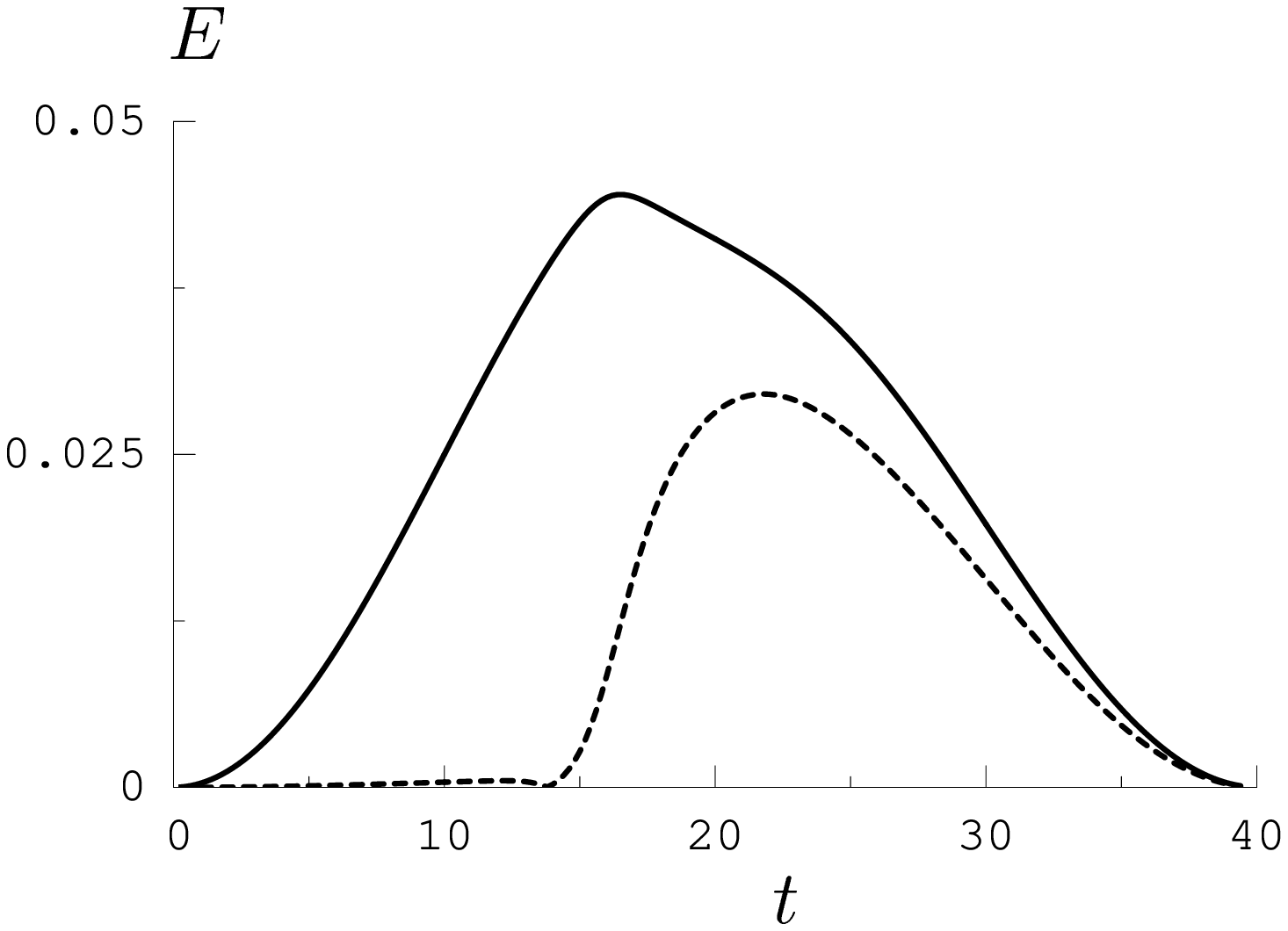} &
\epsfxsize5cm\epsffile{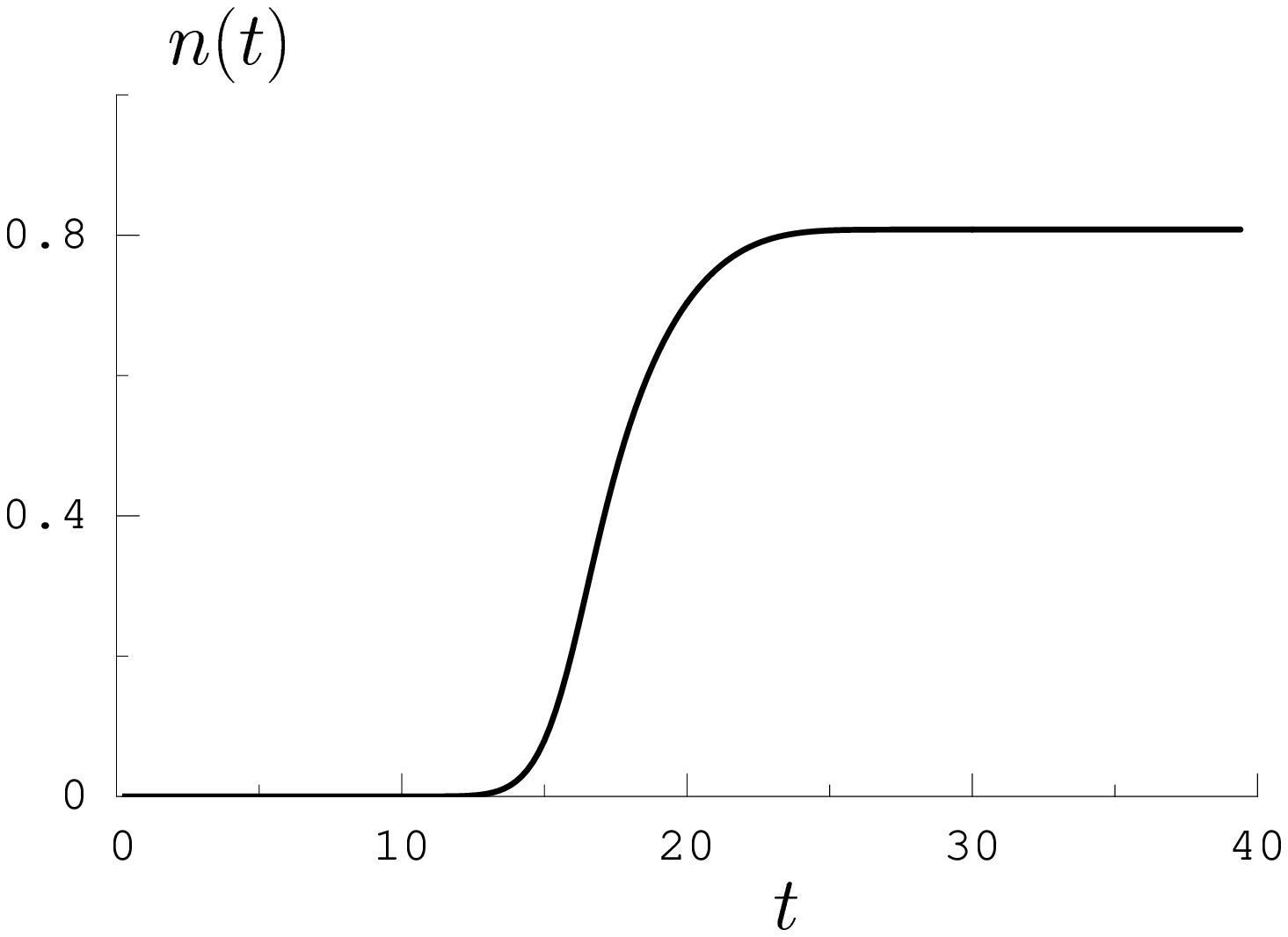} &
\epsfxsize5cm\epsffile{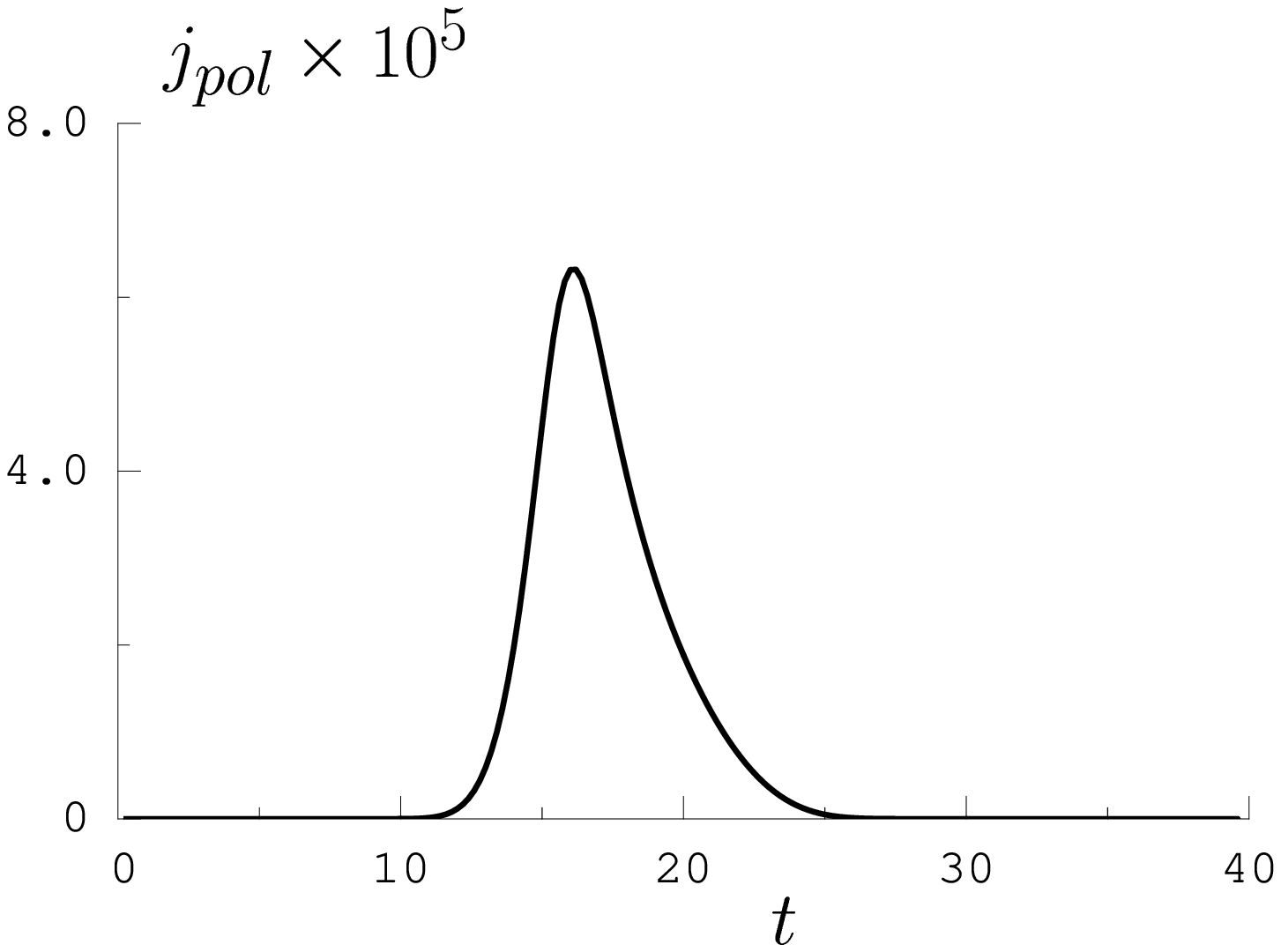}\\
\epsfxsize5cm\epsffile{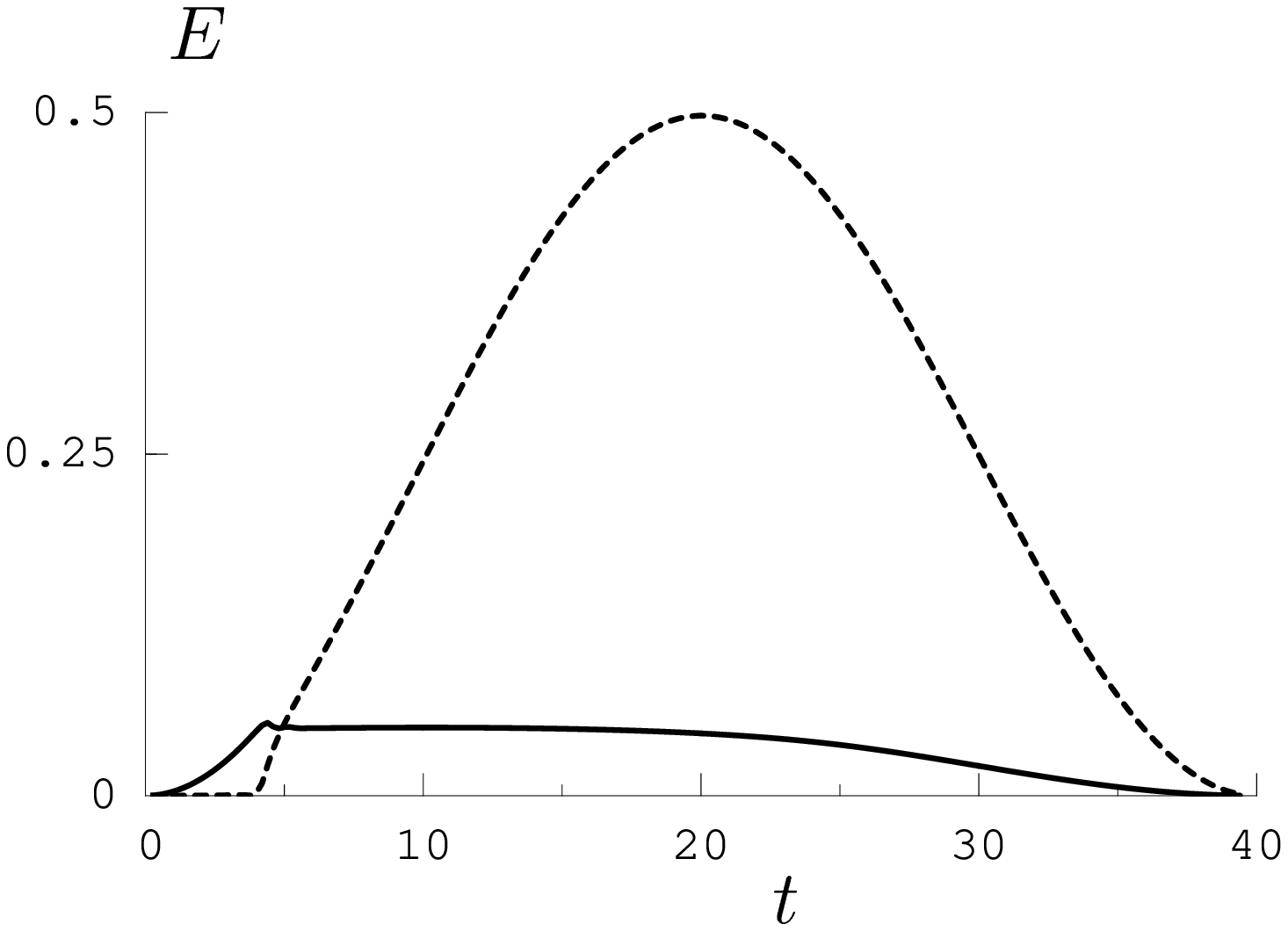} &
\epsfxsize5cm\epsffile{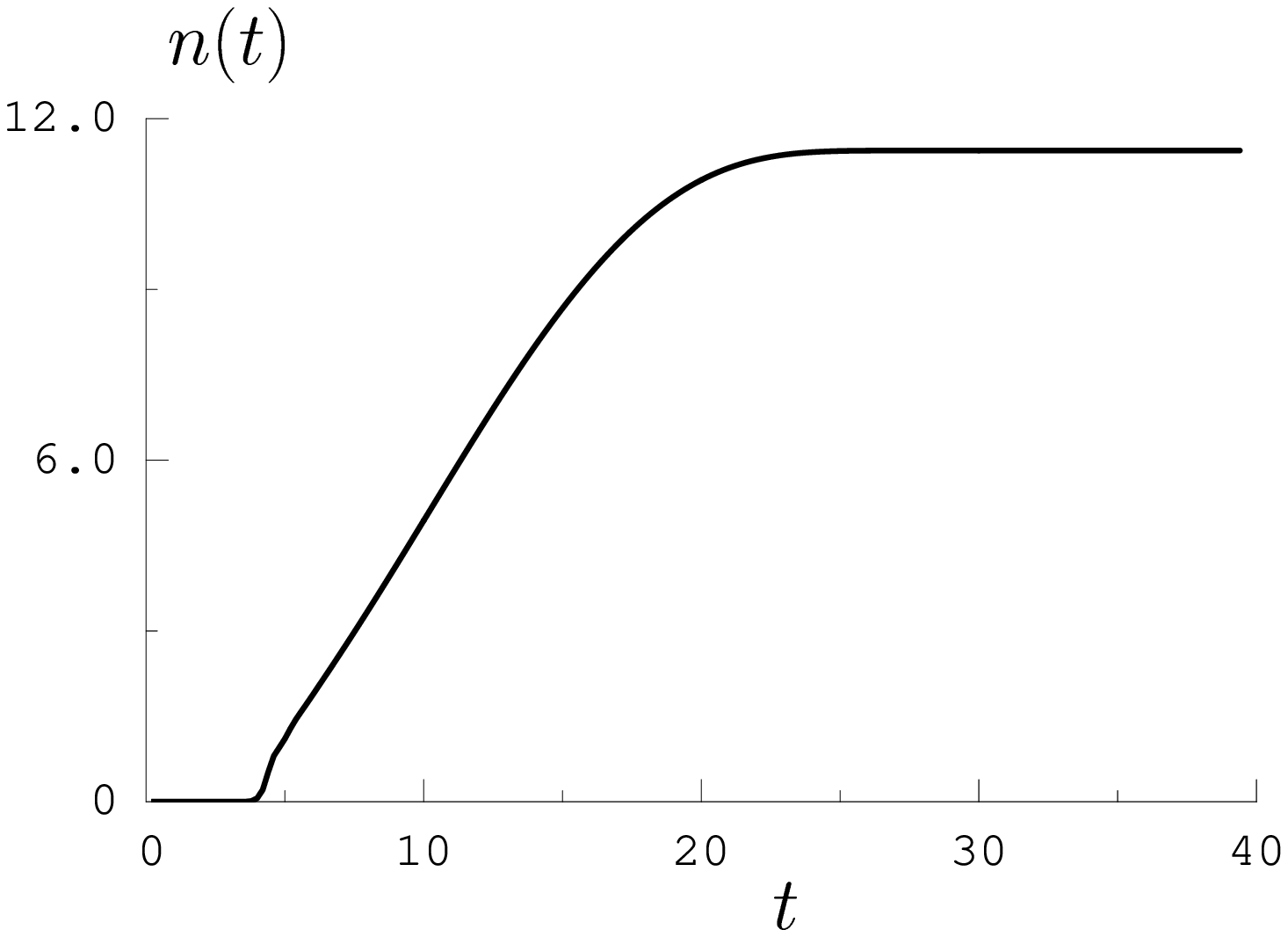} &
\epsfxsize5cm\epsffile{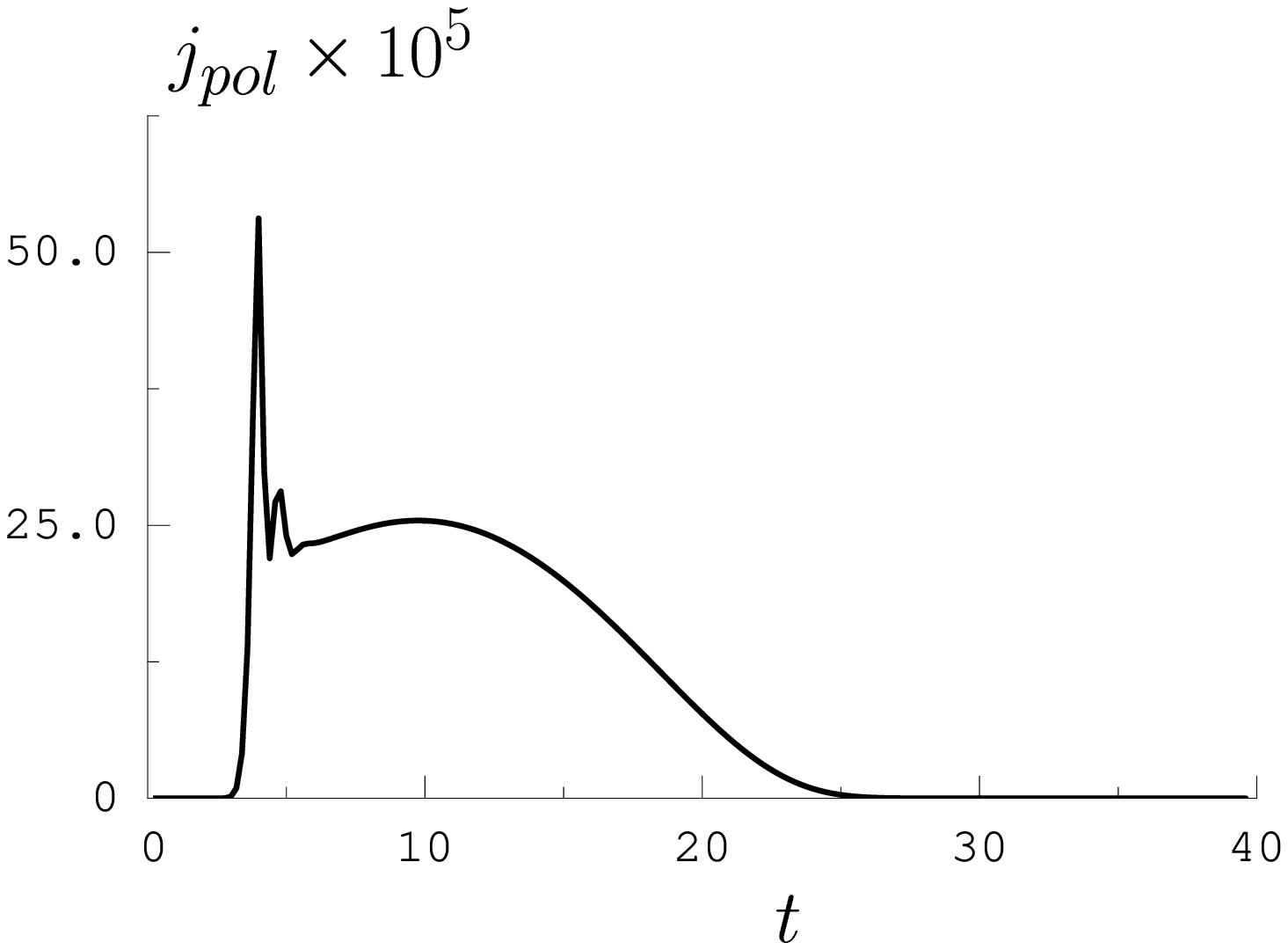}\\ a) & b) & c)\\
\end{tabular} \caption{Evolution of the
reflected (dotted curve) and transmitted pulse electric field (a),
plasma density (b) and polarization current (c) for initial pulse of
the form (\ref{sin}). Time is measured in units of wave period
($2\pi/\omega$), field is measured in units of $mc\omega/e$, density
is measured in units of critical density, $n_{cr}$, and currents are
dimensionless. The parameter $\epsilon_p K_0=4\times 10^3$,
$x=0$.The first row corresponds to an initial pulse amplitude equal
to $\eta_0^{max}=0.05$, the second row to $\eta_0^{max}=0.5$.}
\end{figure}

In the case of $\eta_0^{max}=0.2$ (Fig. 1, second row) the situation
is different. Since the initial pulse intensity is high enough to
produce a plasma with a density 4 times larger than critical by
means of foil ionization, the reflectivity of the foil tends to
unity, as shown in Fig. 1a the main pulse is almost completely
reflected. However we should note that the intensity of the pedestal
is not enough to ionize the foil, that is why it is completely
transmitted through the foil. When the main pulse arrives at the
foil it quickly ionizes the foil, it can be clearly seen in Fig 1b,
second row, where the plasma density evolution is presented. In this
case the reflected pulse is pedestal-free. So through the
interaction of the intense electromagnetic pulse with rapidly
ionized foil the reflected pulse is almost the same as the initial
one apart from the fact that the pedestal is greatly reduced in the
reflected pulse. We should also note that the reflected pulse has a
steeper front than the initial one and its tail is of the same
(half)width as the initial one (see Fig. 1a). These features are
qualitatively expected from the analysis of the simplified wave
equation carried out in the previous section.

In order to illustrate the contribution of the polarization current
to the evolution of the reflected and transmitted pulses, we present
Fig. 1c. As it was pointed out above, the expected polarization
current contribution should be rather small, since it is
proportional to the ionization potential. However the polarization
current is also proportional to the ionization rate, so it can be
important in the first moments of intense ionization. It can be seen
from Fig. 1c that the polarization current is small compared to
conduction one (to estimate the amplitude of the conduction current,
one can multiply electric field amplitude by the number of
particles), but it is non-zero in the region where the conduction
current is far from its maximal value, and thus its contribution can
be of the order of conduction current contribution.

In Fig. 2, where the results of numerical solution of Eq.
(\ref{field}) for initial pulse (\ref{sin}) are presented, maximal
electric field is $\eta_0^{max}=0.05$ (first row) and
$\eta_0^{max}=0.5$ (second row). The amplitudes of the reflected and
transmitted pulses are shown in Fig. 2a. The evolution of plasma
density can be seen in Fig. 2b, and the behavior of polarization
current in Fig. 2c. In the case of the initial pulse maximal
amplitude of $\eta_0^{max}=0.05$ (first row), the reflected and
transmitted pulses are roughly of the same order, but the reflected
pulse has a steeper front, since the reflectivity of the foil
depends on the density of plasma, which evolution is shown in Fig.
2b. With increasing of the initial pulse amplitude the situation
changes, in the case of $\eta_0^{max}=0.5$ (second row) the
transmitted pulse is much smaller than the reflected, which almost
coincides with the initial one. Since the amplitude of the initial
pulse is high enough to create a significant amount of plasma even
in the first moments of the pulse interaction with the foil, the
initial pulse gets reflected. Here, analogous to the case of the
pulse with the pedestal in interaction with the foil, the reflected
pulse has the characteristic temporal profile. Its front is steeper
than its tail, while the latter (half)width is the same as the
(half)width of the initial pulse. The properties of the polarization
current are the same as in the previous case of pulse with pedestal.
The absolute value of the polarization current is much smaller than
that of the conduction current. However the polarization current has
a maximum in the region where the conduction current is far from
reaching its maximal value.

\begin{figure}[ht]
\epsfxsize6cm\epsffile{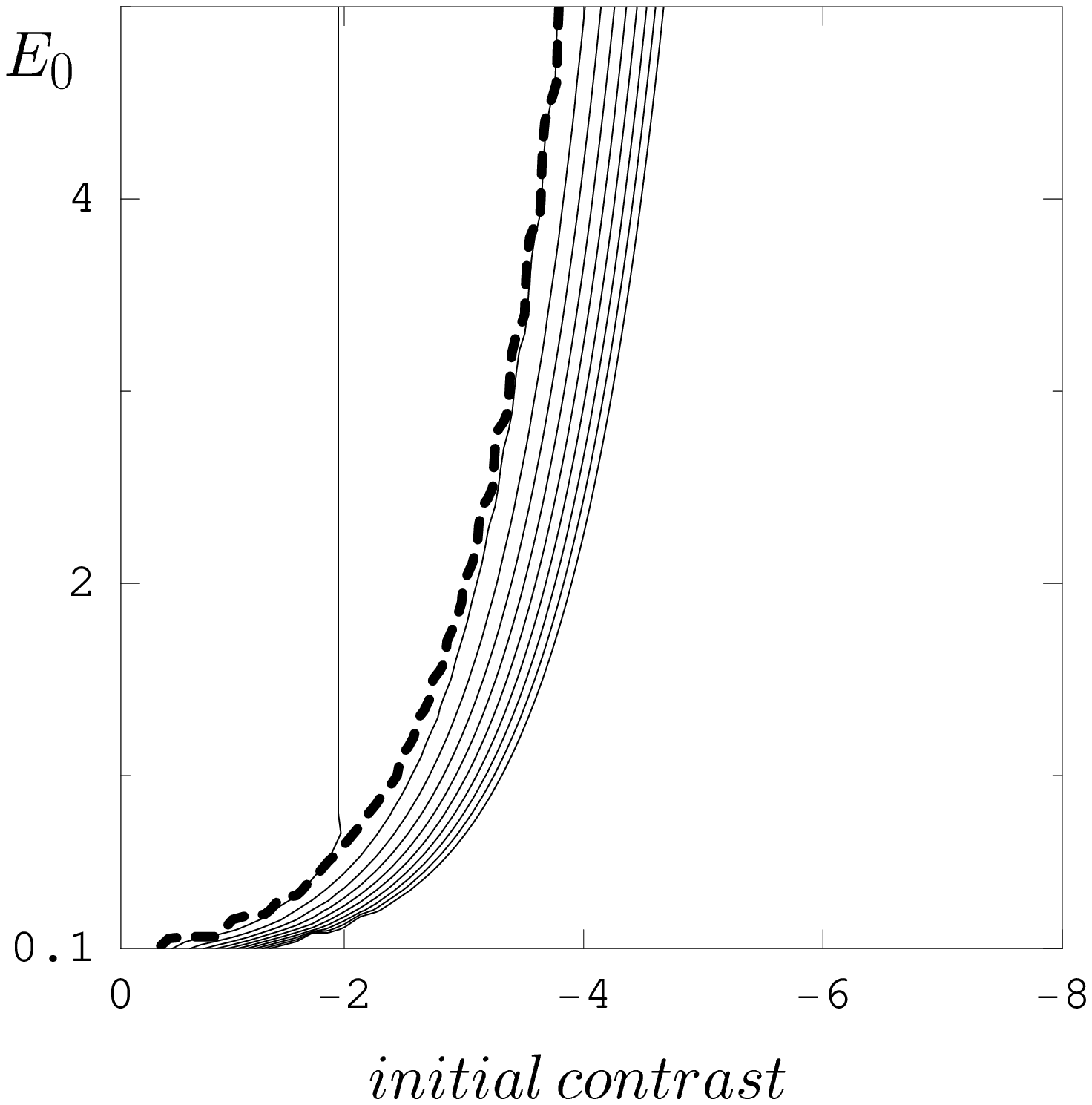} \caption{Dependence of the
reflected pulse contrast (the ratio of the pedestal intensity to the
main pulse intensity) on the initial pulse contrast (horizontal
axis, logarithmic scale) and main pulse dimensionless amplitude
(vertical axis, the field is measured in units of $mc\omega/e$). The
contours represent the values $10^{-2 k}$, $k=1..10$. The dashed
curve depicts values where the resulting contrast is equal to the
initial one.To the right of this curve lie the values of the
improved contrast.}
\end{figure}

Let us consider the case of the pulse with pedestal once more but
from the point of view of obtaining high contrast pulses. In order
to get a high contrast pulse the pedestal should be transmitted
through the foil and the safe-reflection should be triggered by the
main pulse. In Fig. 3 we present a contour plot of the reflected
pulse resulted contrast dependence on the initial pulse contrast and
intensity. It is calculated as the ratio of the reflected pulse
pedestal instantaneous intensity at the point where the pedestal and
the main pulse envelopes intersect and the reflected pulse maximum
intensity. The contours correspond to the values $10^{2k}$,
$k=1..10$ of the resulted contrast. We see that for pulses with the
low initial contrast there is no way to improve it by the reflection
on the rapidly ionized foil. It is due to the fact that the
intensity of the pedestal is high enough to ionize the foil giving
rise to the reflection of the incoming radiation. When the initial
contrast improves there appears a region (to the right of the dashed
line in Fig. 3) of initial pulse intensity values where the
resulting contrast of the reflected pulse is better than that of the
initial pulse. With the further improvement of initial pulse
contrast this region enlarges, and for initial contrast better than
$10^{-5}$, this region covers all the range of considered amplitudes
of the initial pulse. So in order to improve the laser pulse
contrast in the interaction with ionizing foil the initial pulse
contrast should be high enough not to allow the prepulse trigger the
foil ionization-induced transition to opacity.

\section{High order harmonic generation by the linearly polarized pulse.}

Since a circularly polarized pulse does not generate high order
harmonics when it ionizes the foil, we consider the ionization of a
foil by a linearly polarized pulse. In this case all the formulae of
sections 2 and 3 are valid, the only thing that should be taken into
account is that in this case the electrons emerge from under their
barriers with a momentum distribution that has its maximum at zero.
However this distribution is very narrow \cite{Popov}, its width is
much less than the typical momentum that an electron acquires in the
electric field. Due to this fact the initial momentum distribution
can be approximated by Dirac delta function in the expression for
the source term. In this case the equation (\ref{field}) should be
modified, and it becomes
\begin{equation}\label{field_L}
\eta=\eta_0-\epsilon_p\int\limits_0^{t-|x|/c}\dot{n}(t^\prime)
\left[a(t)-a(t^\prime)\right]dt^\prime
-\epsilon_p\frac{I}{mc^2}\frac{\dot{n}(t)}{\eta}.
\end{equation}
In what follows we present the results of numerical solution of
equation (\ref{field_L}) for the same initial pulses as in the
preceding section. In Figs. 4 and 5 the form of the reflected and
transmitted pulses, plasma density evolution and the spectrum of the
reflected pulses are shown for the initial pulse shapes
(\ref{pedestal}) and (\ref{sin}) with the same amplitude and
duration as in Figs. 1 and 2. The behavior of the reflected and
transmitted pulses is almost the same as in the case of the
circularly polarized initial pulse as can be seen from Figs. 4a and
5a. However the evolution of the plasma density is slightly
different. In order to illustrate this we present Figs. 4b and 5b,
where a step-like function is shown, its behavior is similar to that
in the case of circular polarization apart from the step-like
feature. This property of the density evolution is due to the
ionization. Since the ionization rate is determined by the absolute
value of the electric field, in the case of the linear polarization
it has two maxima per period of the light wave. The step-like
function just demonstrates that the ionization occurs when the
electric field amplitude reaches its maximum, while there is almost
no ionization in between.

\begin{figure}[ht]
\begin{tabular}{ccc}
\epsfxsize5cm\epsffile{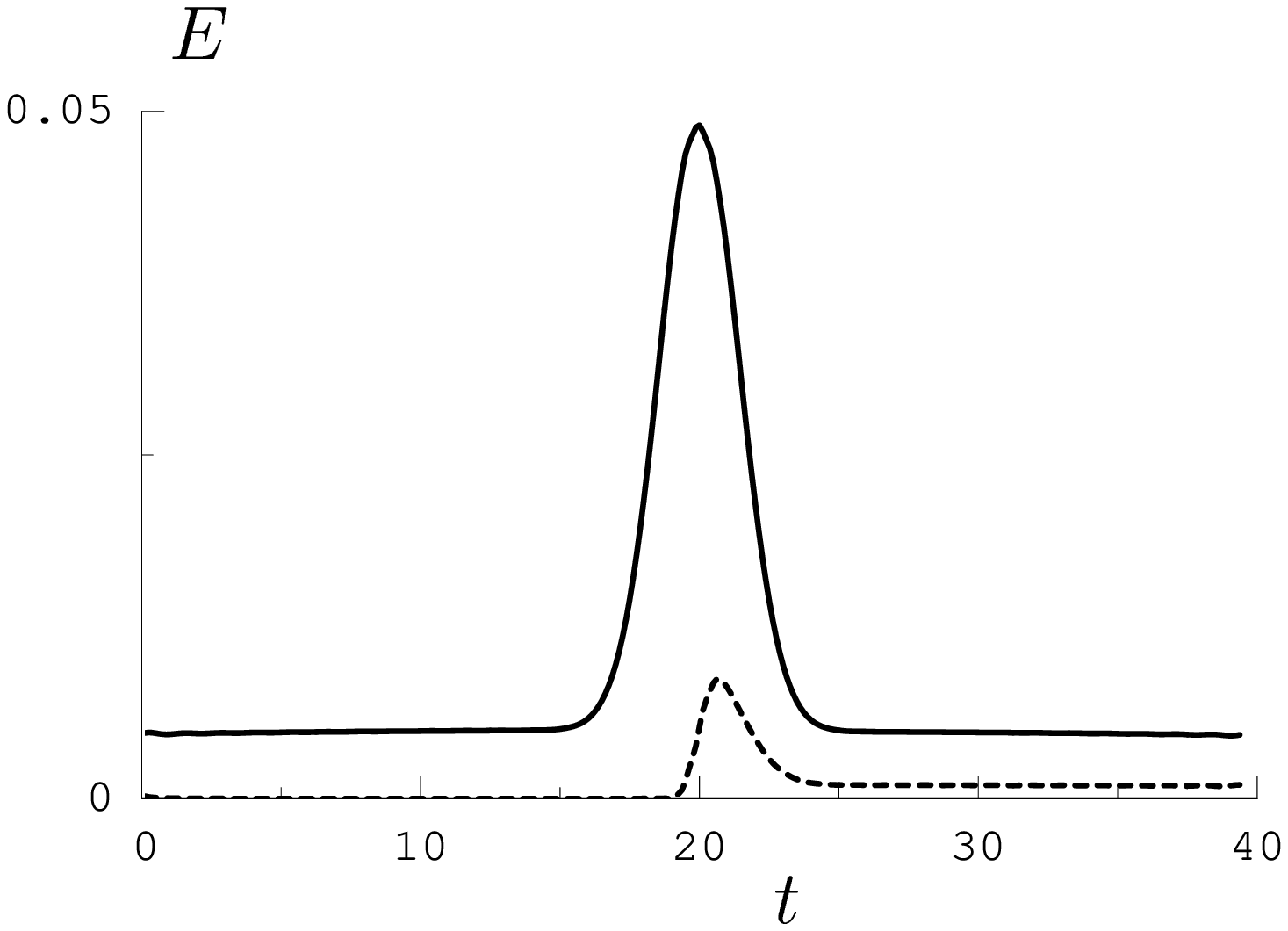} &
\epsfxsize5cm\epsffile{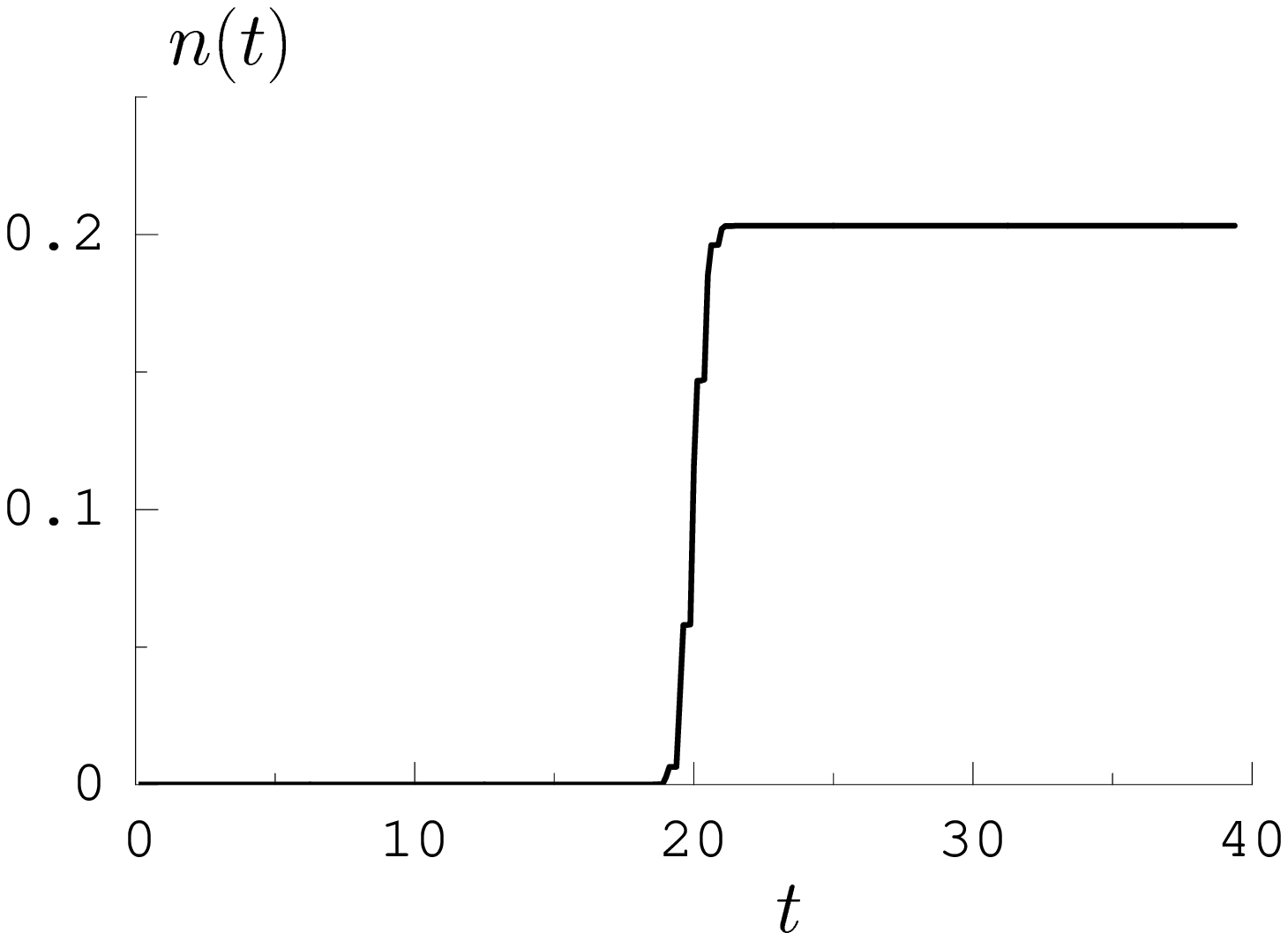} &
\epsfxsize5cm\epsffile{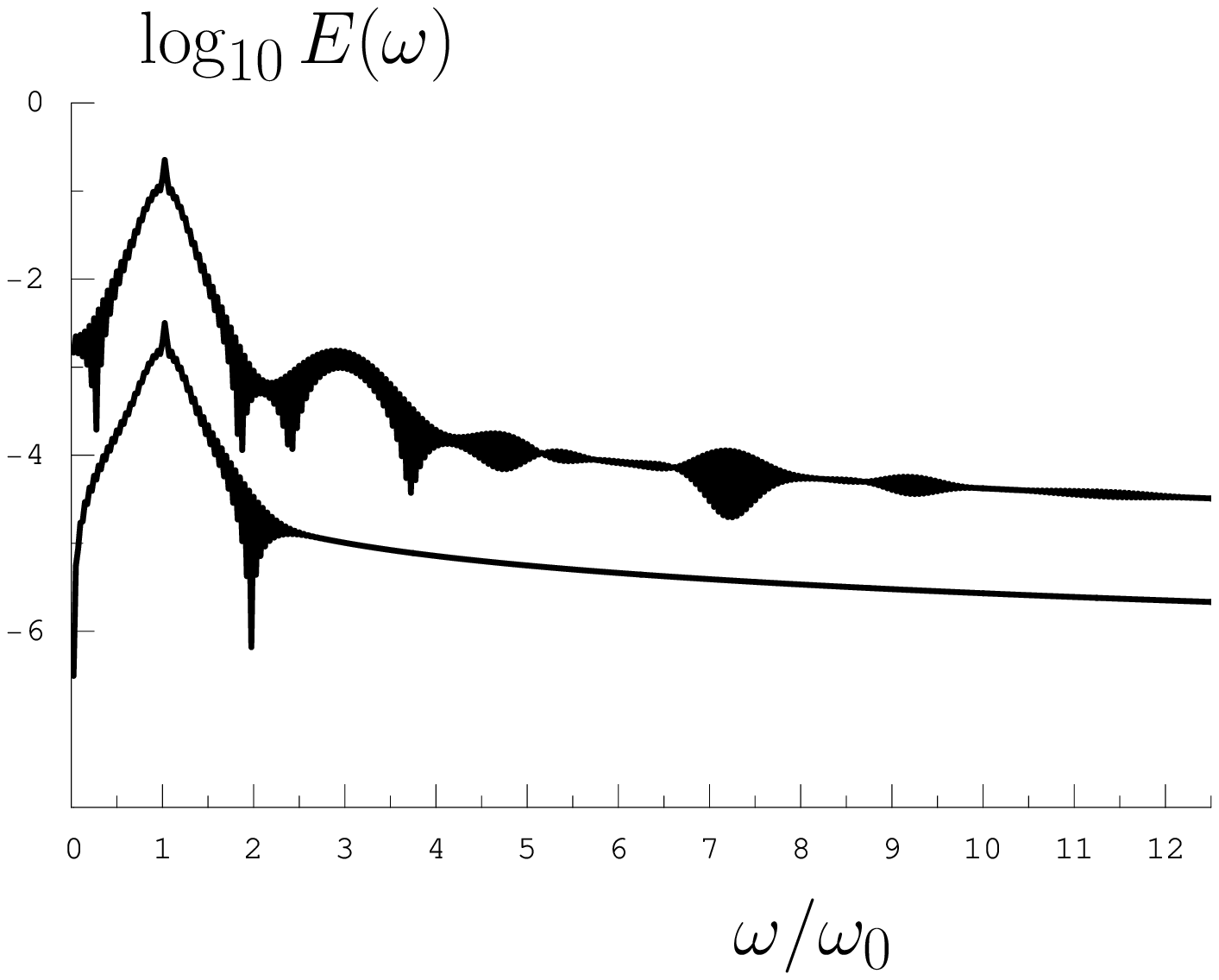}\\
\epsfxsize5cm\epsffile{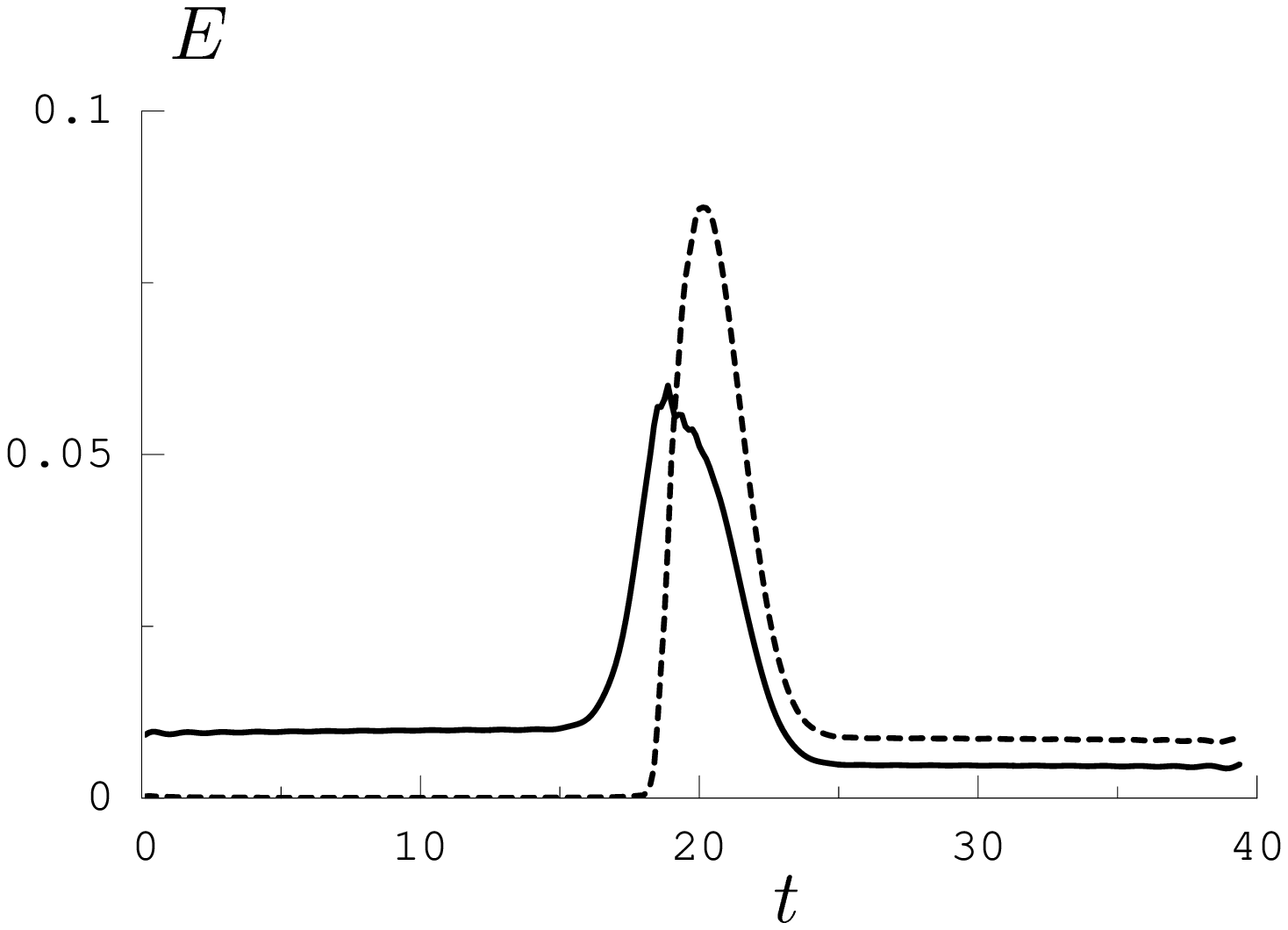} &
\epsfxsize5cm\epsffile{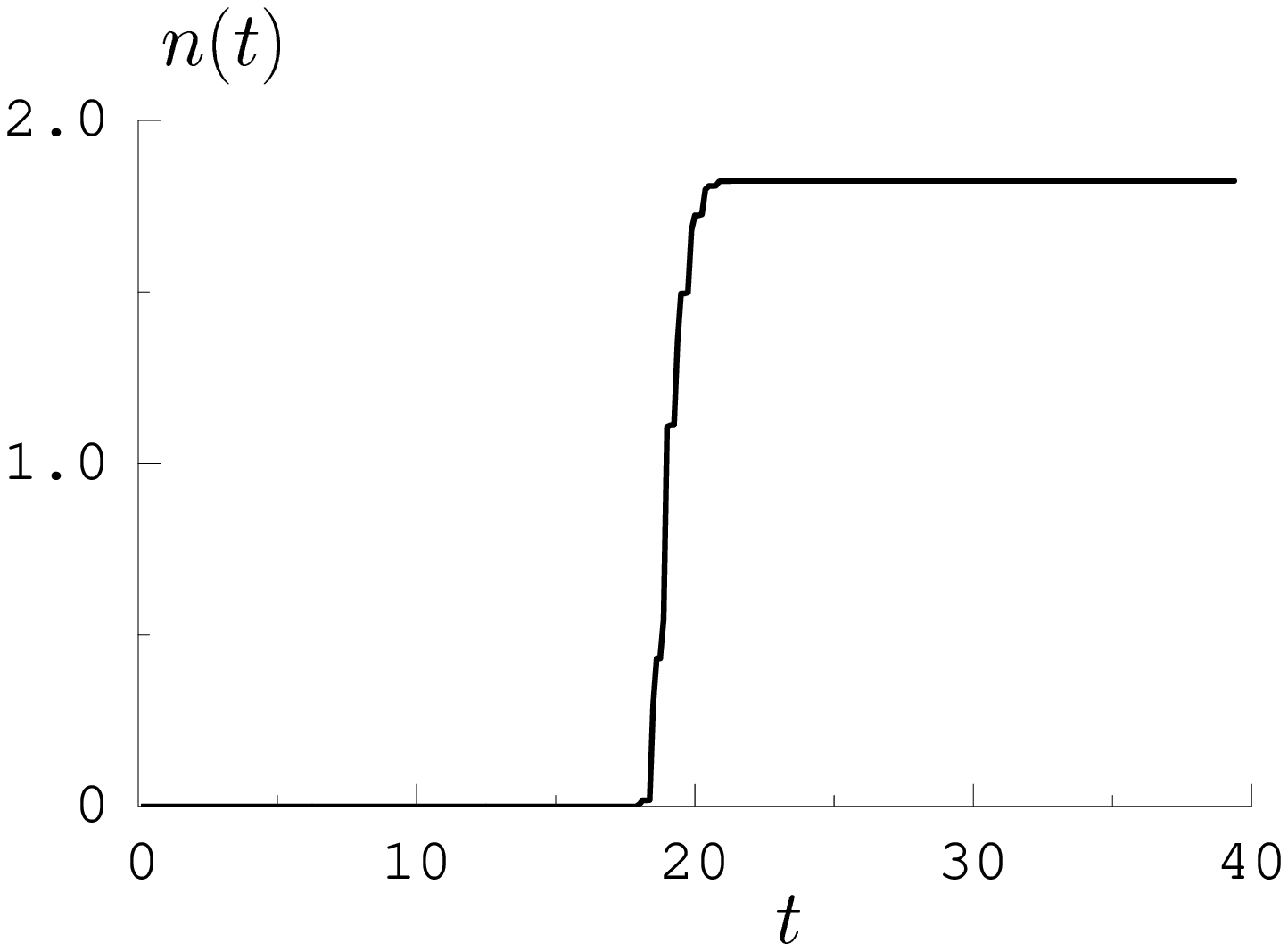} &
\epsfxsize5cm\epsffile{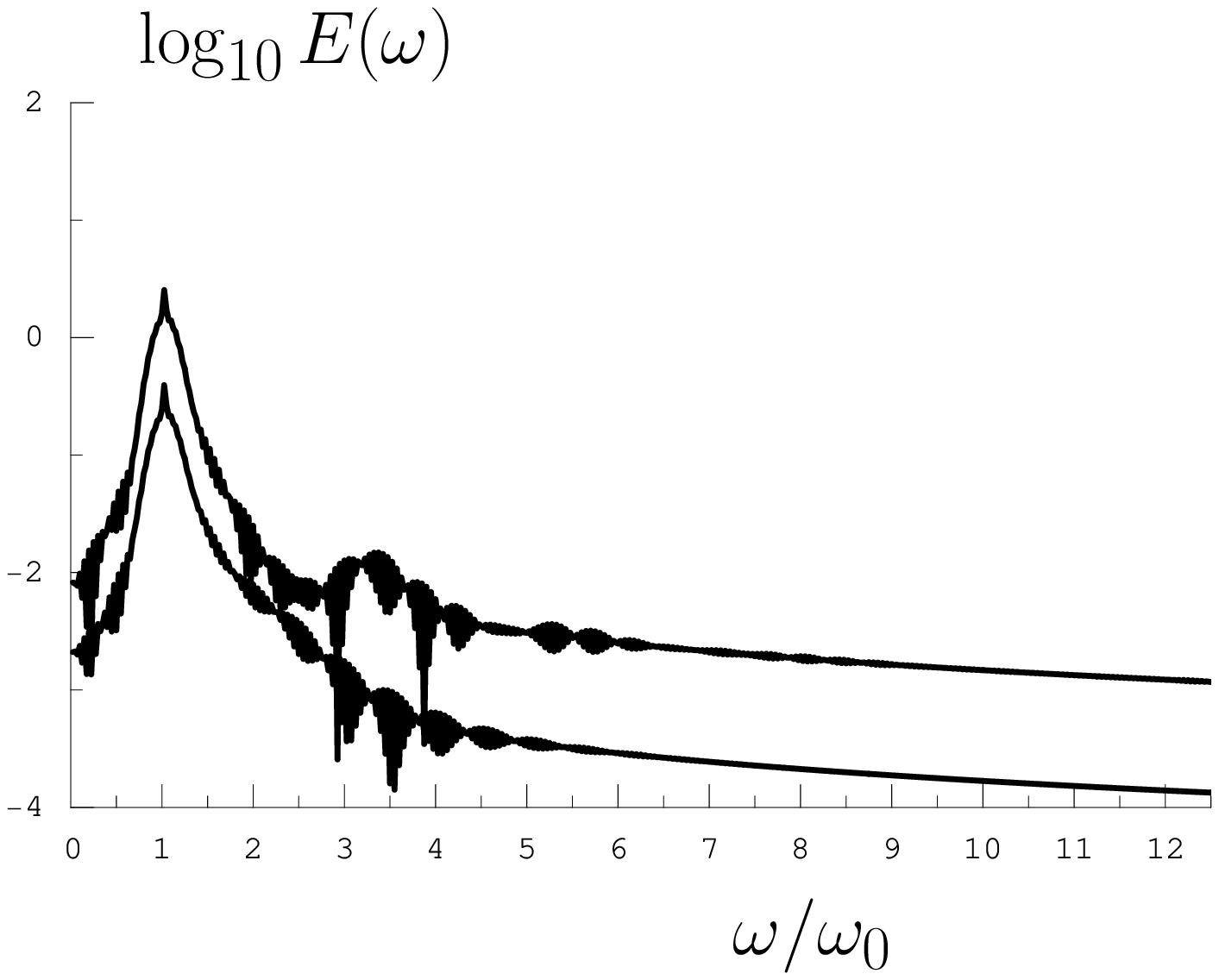}\\ a) & b) & c)\\
\end{tabular}
\caption{Evolution of reflected (dotted curve) and transmitted
electric field envelopes (a) and plasma density (b) in the case of
linear polarization, the spectra of the reflected pulses in the case
of linear (upper curve) and circular (lower curve) polarizations for
the same pulse energy (c) for initial form of the pulse
(\ref{pedestal}). Time is measured in units of wave period
($2\pi/\omega$), field is measured in units of $mc\omega/e$ and
density is measured in units of critical density, $n_{cr}$. The
parameter $\epsilon_p K_0=4\times 10^3$. The first row corresponds
to the initial pulse amplitude equal to $\eta_0^{max}=0.05$, the
second row to $\eta_0^{max}=0.1$.}
\end{figure}

In Figs. 4c and 5c the spectra of the reflected pulses are
presented. We see that due to the nonlinear dependence of the
ionization rate on the absolute value of the field harmonic
generation occurs. Since we consider only normal incidence of the
laser pulse on the foil the electric field is parallel to the foil,
and that is why only odd harmonics are generated as it can be seen
from Figs. 4c and 5c. In order to compare the cases of linear and
circular polarizations we also present in Figs. 4c and 5c the
spectra of circularly polarized reflected pulses, that have the same
energy as the linearly polarized ones. Here we can clearly see that
there is no high order harmonic generation in the case of circularly
polarized pulses. Whereas the modification of the spectrum that is
due to the ionization blueshift, that was mentioned in Section IV,
manifests itself in both cases. However this effect is weak when the
plasma mirror works in the condition close to the optimal regime as
the transparency-opacity transition is abrupt, since the interaction
length is very small, of the order of the skin depth.

\begin{figure}[ht]
\begin{tabular}{ccc}
\epsfxsize5cm\epsffile{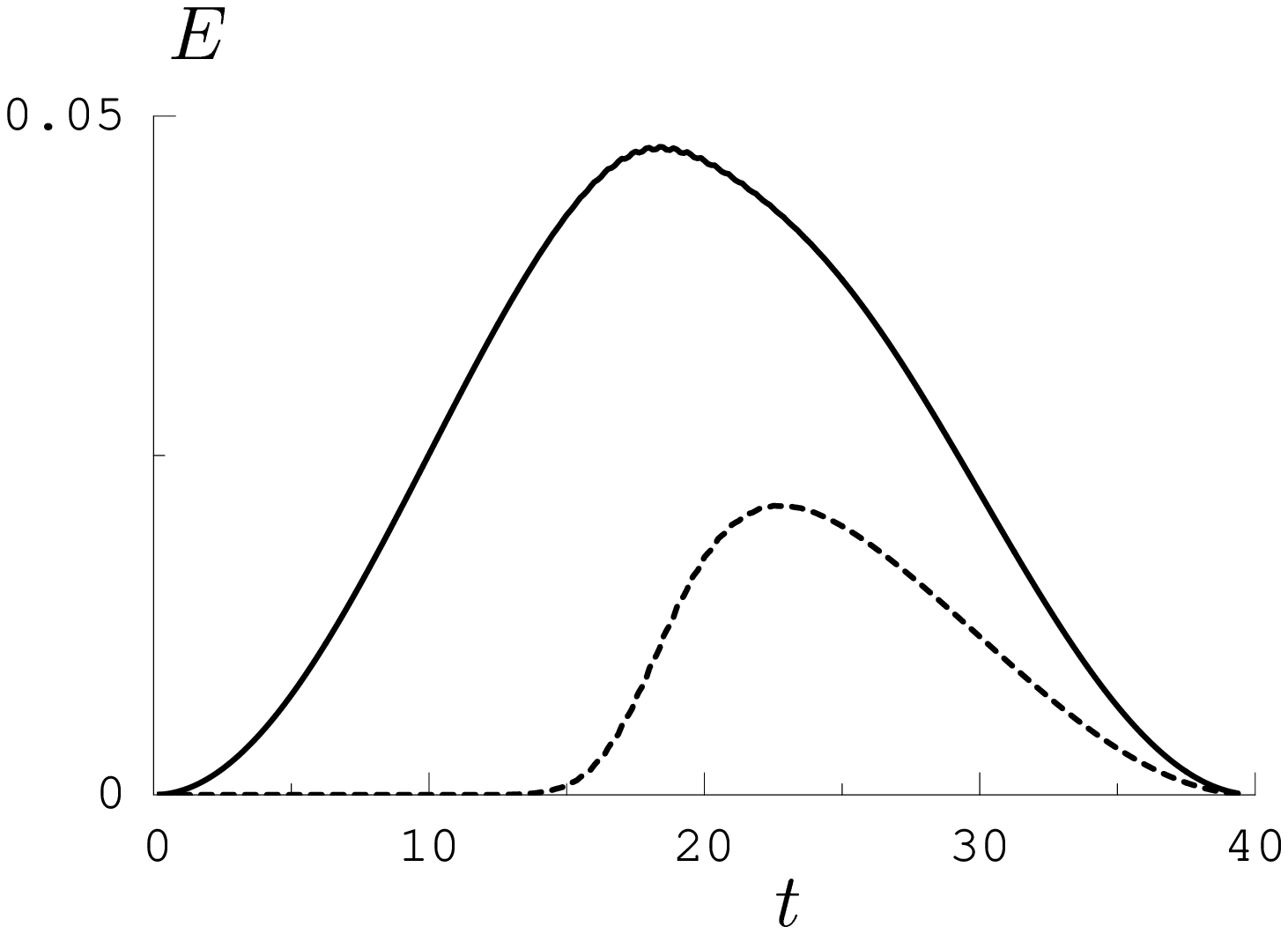} &
\epsfxsize5cm\epsffile{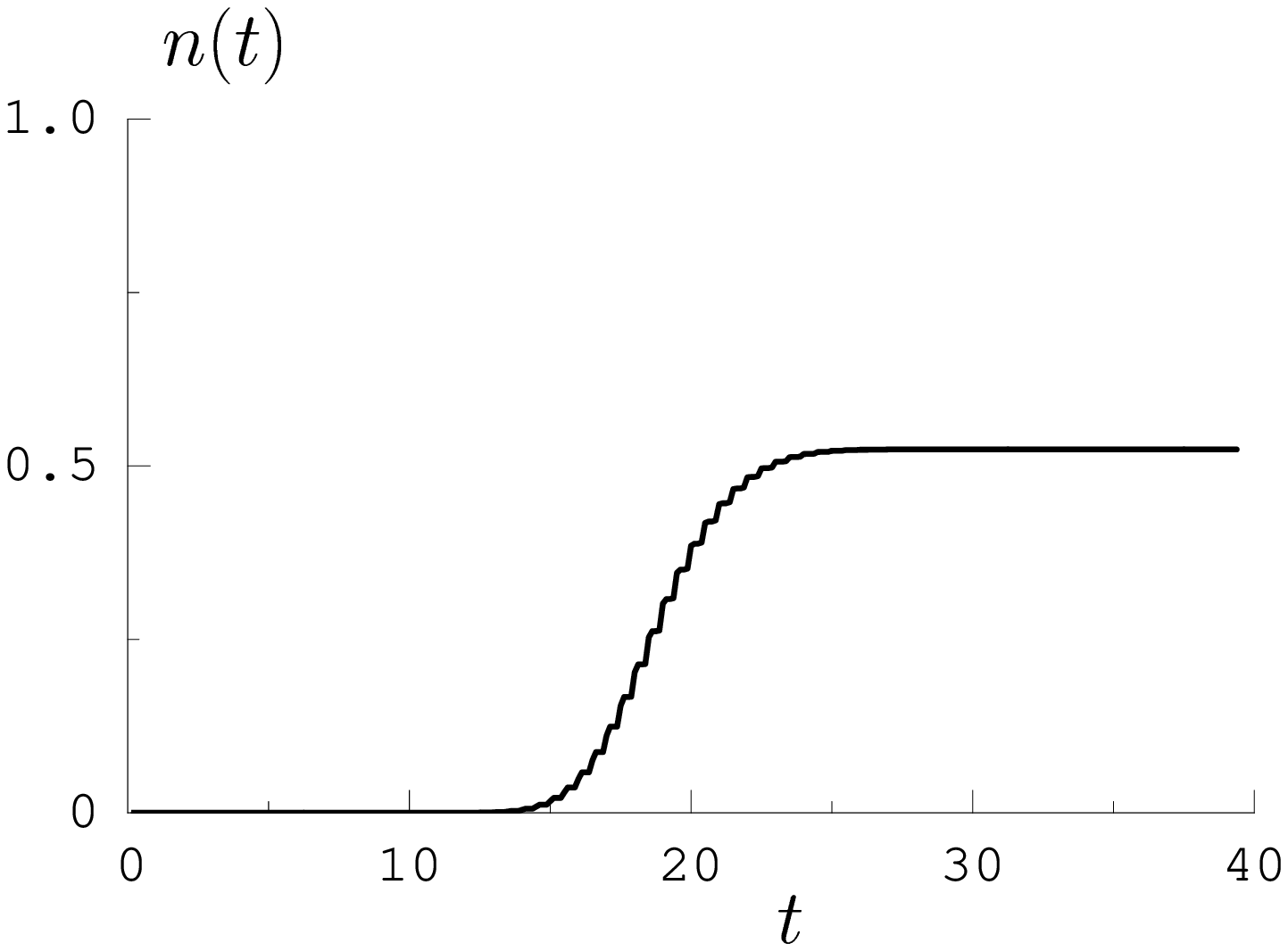} &
\epsfxsize5cm\epsffile{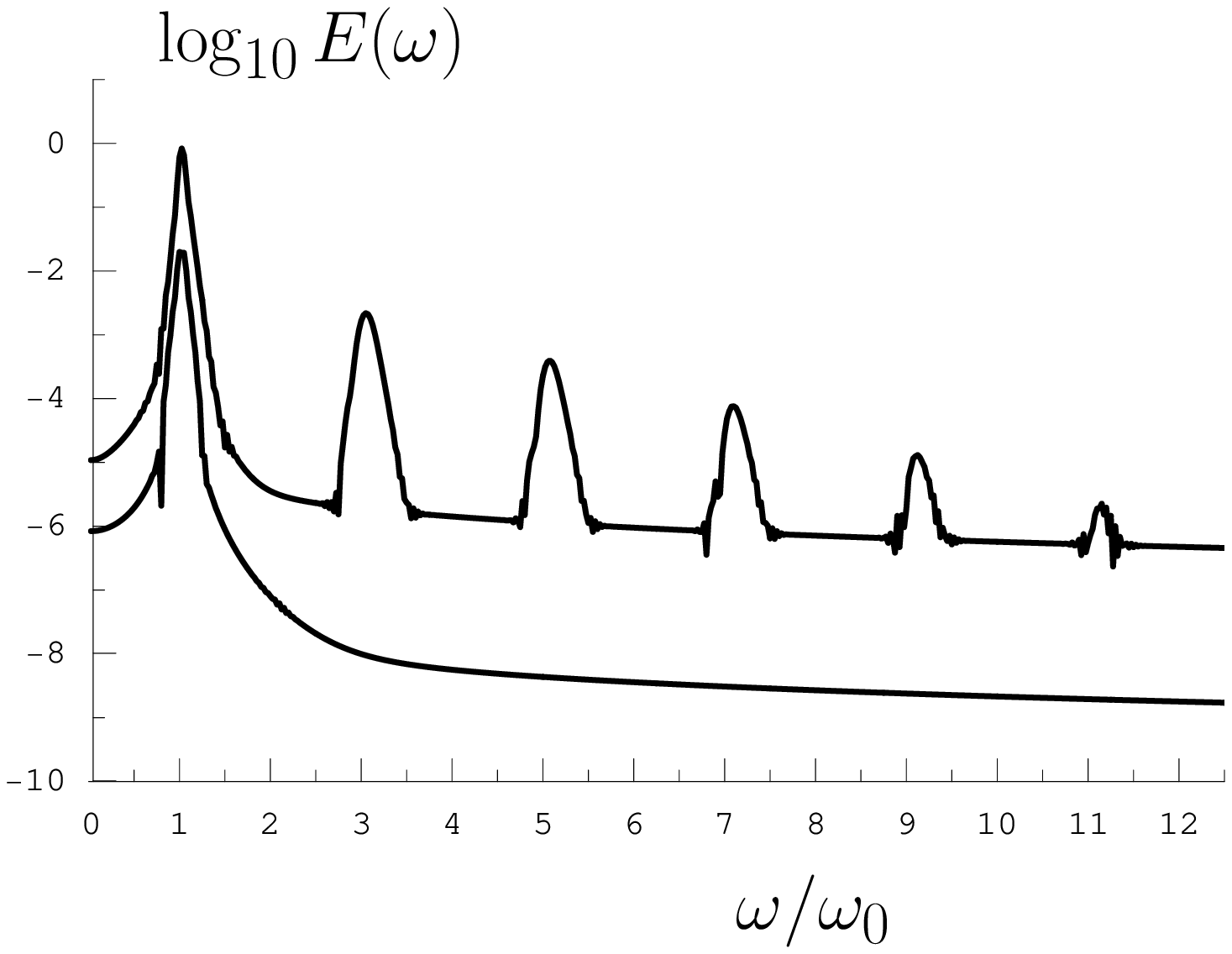}\\
\epsfxsize5cm\epsffile{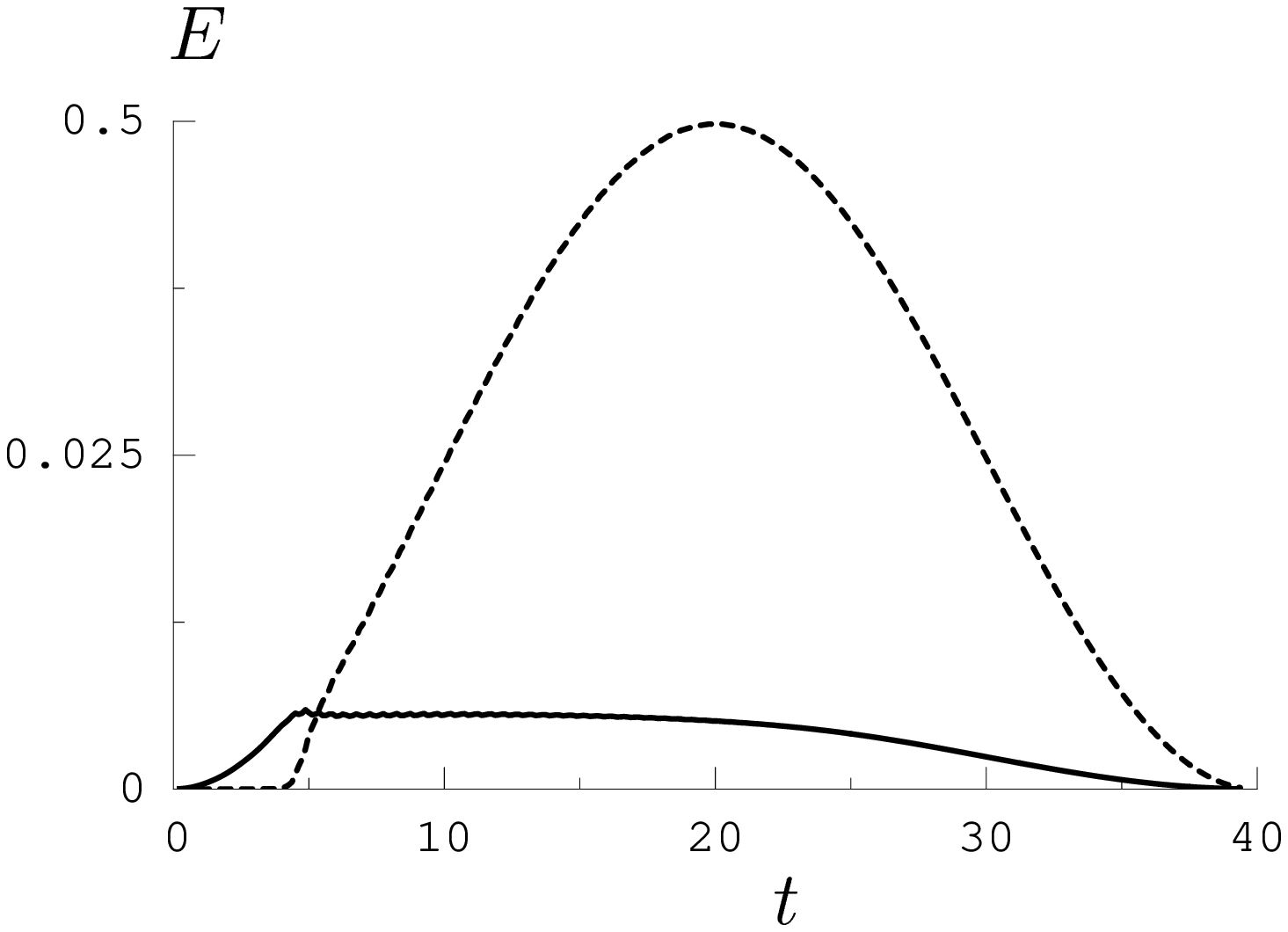} &
\epsfxsize5cm\epsffile{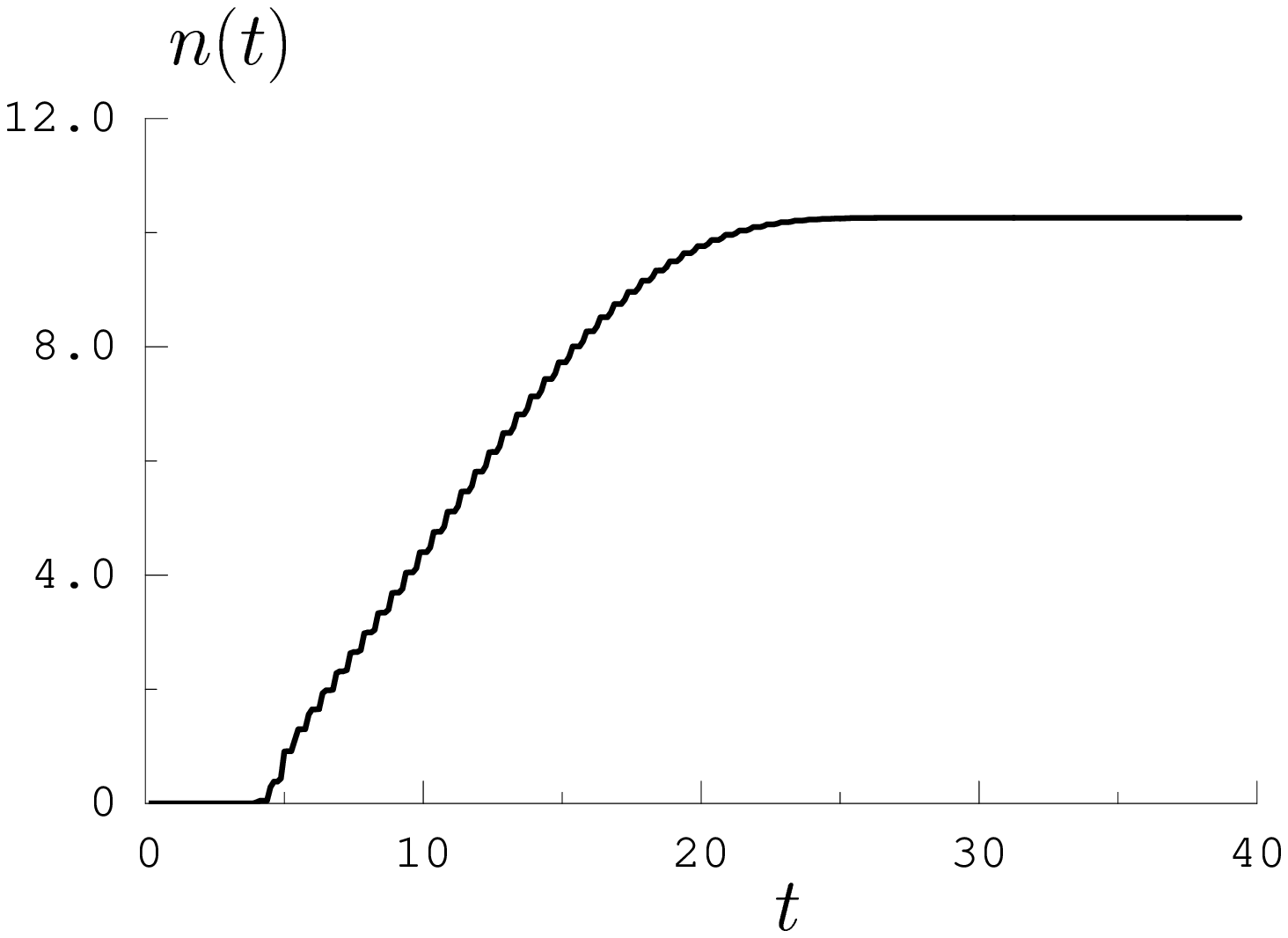} &
\epsfxsize5cm\epsffile{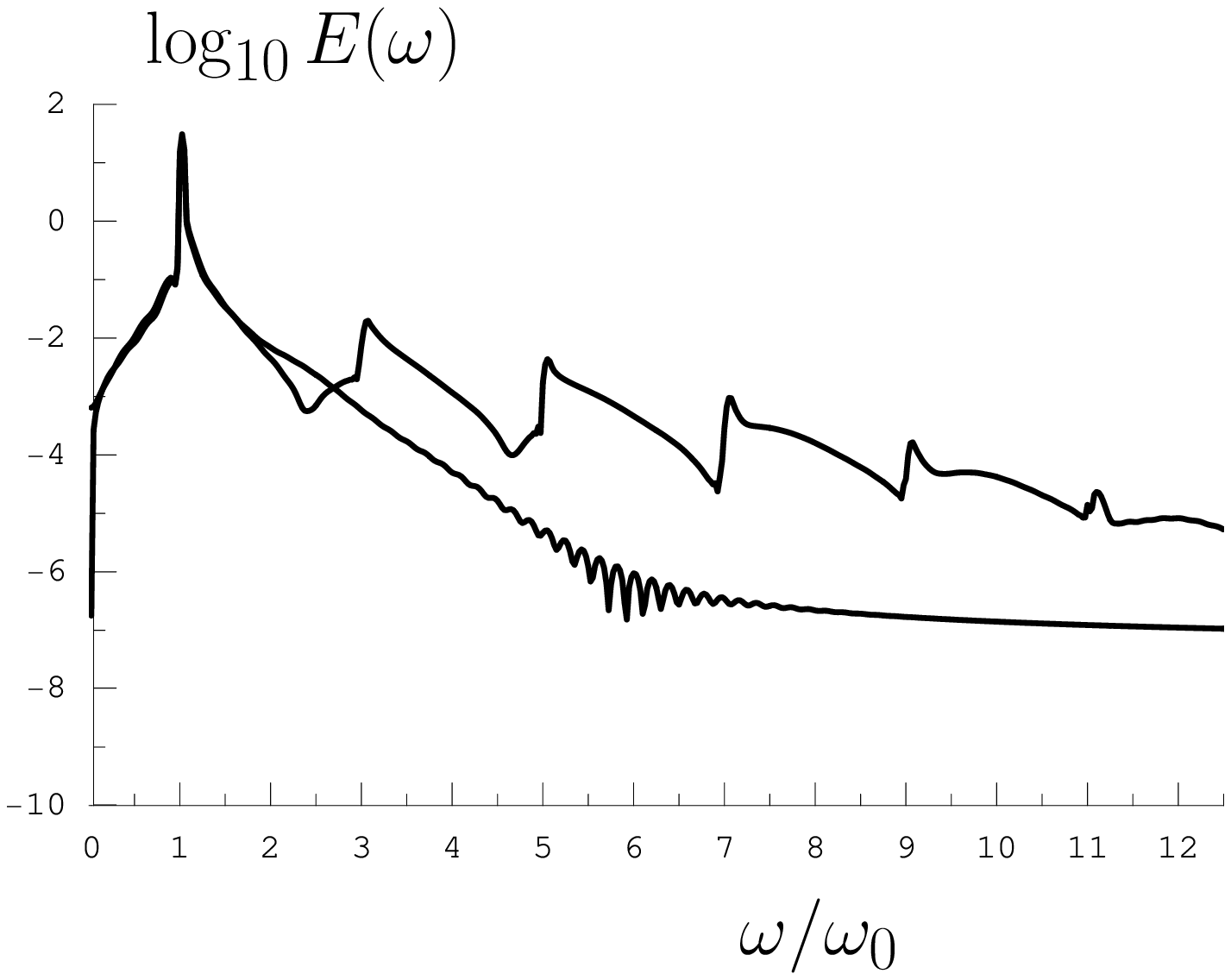}\\ a) & b) & c)\\
\end{tabular}
\caption{Evolution of reflected (dotted curve) and transmitted
electric field envelopes (a) and plasma density (b) in the case of
linear polarization, the spectra of the reflected pulses in the case
of linear (upper curve) and circular (lower curve) polarizations for
the same pulse energy (c) for the initial form of the pulse
(\ref{sin}). Time is measured in units of wave period
($2\pi/\omega$), field is measured in units of $mc\omega/e$ and
density is measured in units of critical density, $n_{cr}$. The
parameter $\epsilon_p K_0=4\times 10^3$. The first row corresponds
to the initial pulse amplitude equal to $\eta_0^{max}=0.05$, the
second row to $\eta_0^{max}=0.5$.}
\end{figure}

Let us estimate the collective efficiency of high order harmonic
generation. In order to do so we define the parameter $R$:
\begin{equation}
R=\frac{\int\limits_{\omega=2}^{\omega=\infty}E_{ref}^2(\omega)d\omega}
{\int\limits_{\omega=0}^{\omega=\infty}E_{in}^2(\omega)d\omega}.
\end{equation}
This parameter has the meaning of the energy that is contained in
higher harmonics in the units of initial pulse energy. In other
words it measures the efficiency of initial pulse energy
transformation into high frequency radiation.

\begin{figure}[ht]
\begin{tabular}{cc}
\epsfxsize5cm\epsffile{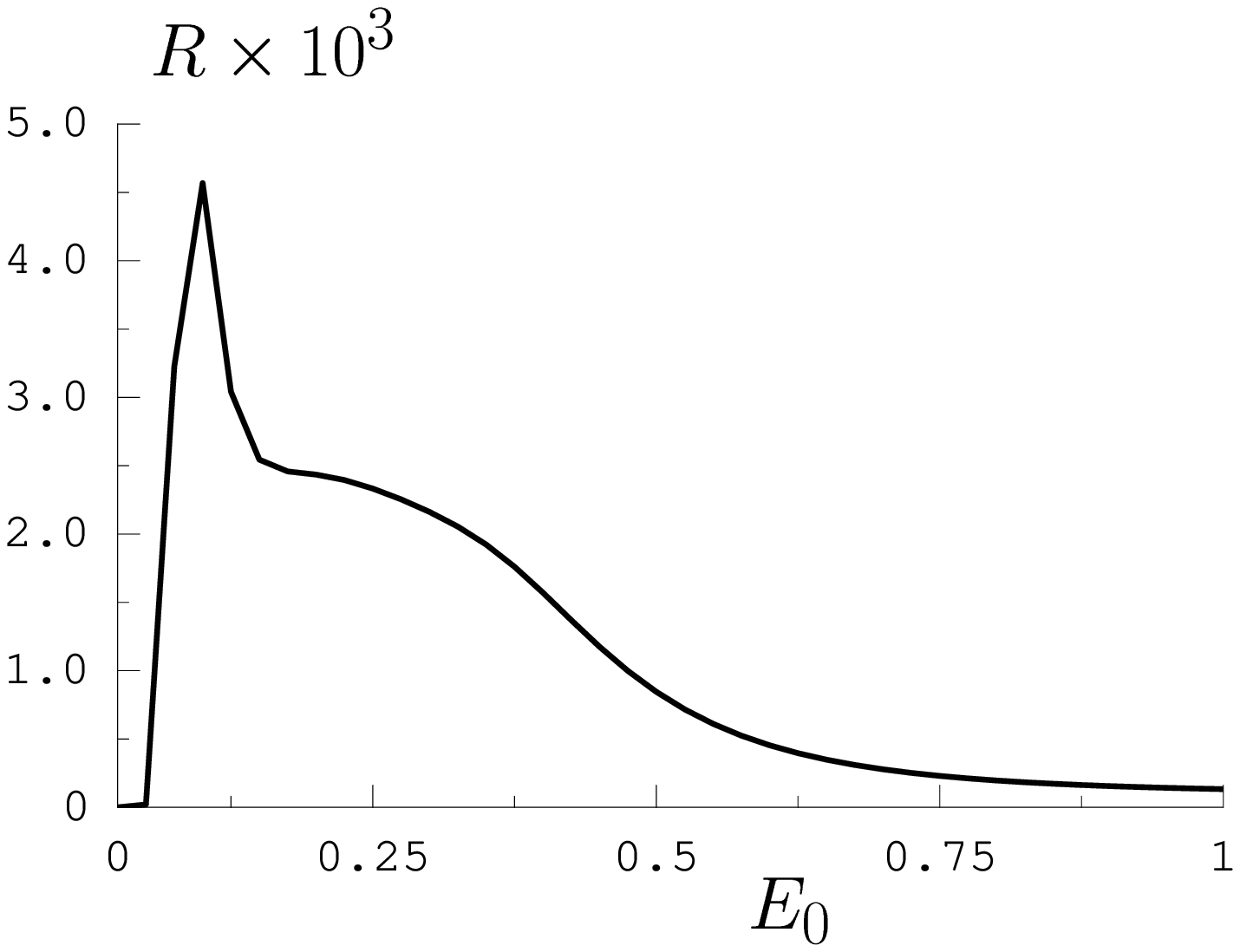} &
\epsfxsize5cm\epsffile{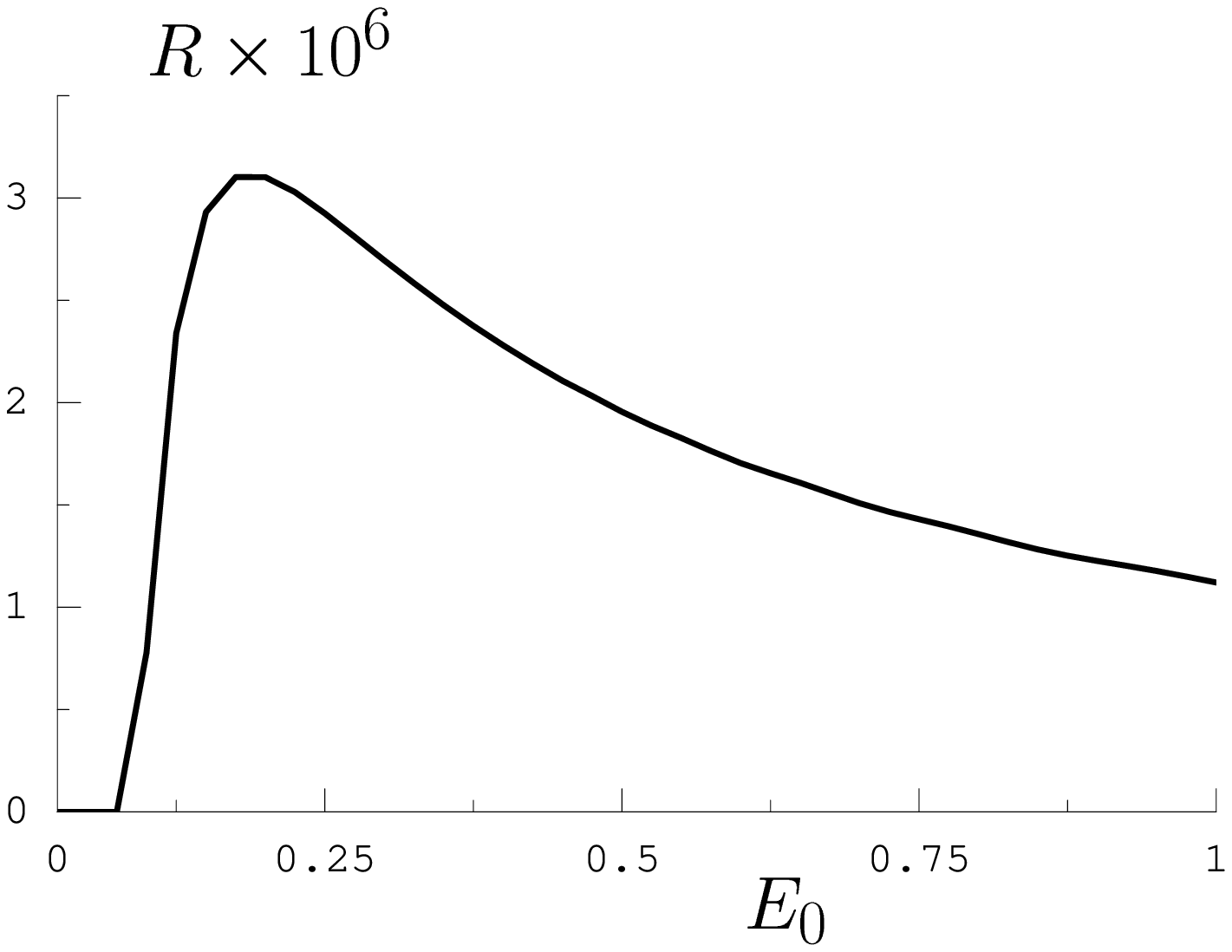}\\
 a) & b) \\
\end{tabular} \caption{Effectiveness of
harmonic generation vs the initial field amplitude for the cases of
initial pulse form (\ref{pedestal}) (a) and (\ref{sin}) (b). The
parameter $\epsilon_p K_0=4\times 10^3$.}
\end{figure}

The dependence of $R$ on the initial pulse amplitude is shown in
Fig. 6 for different forms of the initial pulse: a) gaussian pulse
with pedestal (\ref{pedestal}) and b) pulse with sine squared
envelope (\ref{sin}). The behavior of the curves is almost the same,
apart from the fact that in Fig. 6a the parameter $R$ is measured in
the units of $10^{-3}$, while in Fig. 6b in the units of $10^{-6}$.
For small values of the initial pulse electric field amplitude $R$
is equal to zero. This corresponds to the fact that there is no
ionization for low values of $E_0$. Then at $E_0\approx 0.1$ there
is a maximum of $R$. At this value of the initial field amplitude
the reflected and transmitted pulses are roughly of the same
magnitude. So the high harmonic generation due to foil ionization is
most effective where the duration of ionization is the longest.

\section{Conclusions}\label{CC}

A model for the interaction of a short laser pulse with a thin foil
target undergoing field ionization has been derived. The ionization
dynamics have been described starting from a kinetic equation for
the electrons where the source term includes the ionization rate and
the momentum distribution of freed electrons. Energy conservation
has been included via a polarization current term. Using the
electromagnetic boundary conditions at the foil surface a set of
ordinary differential equations has been derived. Such a compact
model can be used for simple analysis and preliminary parametric
studies of the ``plasma mirror'' properties of the thin foil, such
as prepulse suppression and ultrafast optical shuttering. As an
example the transmission, reflection and shaping of a short laser
pulse with a pedestal have been studied as a function of the initial
pulse contrast and peak intensity. The generation of high harmonics
by ionization has also been studied for a linearly polarized pulse.

Further extension of this compact model will require the inclusion
of multiply ionized species and off-normal incidence laser pulses
that can be achieved by the modification of the source term and the
proper transformation of the reference frame correspondingly. This
will allow the study of the "plasma mirror" operation with respect
to the angle of incidence and different states of polarization.

\acknowledgments

The authors would like to acknowledge fruitful discussions with V.
D. Mur, V. S. Popov and V. Yanovsky. This work was supported by the
National Science Foundation through the Frontiers in Optical and
Coherent Ultrafast Science Center at the University of Michigan and
in part by the Russian Fund for Fundamental Research under project
04-02-17157.

\end{document}